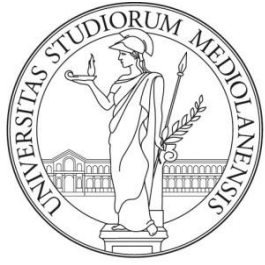

**UNIVERSITÀ DEGLI STUDI DI MILANO**
FACOLTÀ DI SCIENZE MATEMATICHE, FISICHE E NATURALI
DOTTORATO DI RICERCA IN
FISICA, ASTROFISICA E FISICA APPLICATA

# Near Field Speckles: The Optical Theorem Revisited

Settore Scientifico disciplinare FIS/03

**Coordinatore:** Prof. Marco Bersanelli
**Tutore:** Prof. Marzio Giglio
**Cotutore:** Dr. Marco Alberto Carlo Potenza

**Tesi di Dottorato di:**
Sabareesh Kunjipalayam Palaniswamy Velu
**Ciclo XXII**

**Anno Accademico 2008-2009**

கற்க கசடறக் கற்பவை கற்றபின்
நிற்க அதற்குத் தக.

          -- திருவள்ளுவர்

**Lore worth learning, learn flawlessly**
**Live by that learning thoroughly.**

          -- **Thiruvalluvar**

**Palani Vaithiar**
பழனி வைத்தியர்

**Kittammal**
கிட்டம்மாள்

**Nachammal**
நாச்சம்மாள்

# ACKNOWLEDGEMENT

Thanking "THE POWER OF NATURE", the Almighty, for all his blessings, I bring forth my acknowledgement.

First and foremost I would like to express my sincere gratitude to my tutor, *Prof. Marzio Giglio* for his supervision, advice and guidance from the very early stage of this research as well as giving me extraordinary experiences throughout the work. His truly scientist intuition has made him as a constant oasis of ideas and passions in science, which exceptionally inspire and enrich my growth as a student, a researcher and a scientist want to be. I am greatly indebted to him.

I wish to express my profound gratitude and sincerity to my co-tutor *Dr. Marco A. C. Potenza*, for his valuable guidance, suggestions, persistent encouragement and for providing me liberty throughout my doctoral program. I have been benefited by his guidance and always kindly grant me his time even for answering some of my unintelligent questions. Furthermore, he has spent his precious time to read this thesis and has given constructive comments on it.

I'm greatly indebted to *Dr. Mateo Alaimo* for his amicable help in LabVIEW programming. His involvement with his originality has triggered and nourished my intellectual maturity. He has shown me the spirit and motivation towards the carrier and has encouraged me to overcome the difficult tasks of my work.

My special thanks to *Dr. Marina Carpineti* for her valuable suggestions on colloidal aggregation. She has been so good and kind to me and I express my sincere gratitude to her for I had learnt the art of physics from her. I'm really surprised on her efforts to create interest towards science among children.

My thanks to *Prof. Alberto Vailati* for being supportive to my research work.

I extend my profound thanks to our Doctorate School of Physics, Astrophysics and Applied Physics, and to the Director *Prof. Marco Bersanelli* for providing me all needs and has taken prior steps to make my stay and research comfortable. It will not be trustworthy if I did not mention *Prof. Gianpaolo Bellini*, the former Director of the Doctorate School, for he has taken special care and attention especially towards the foreign students and providing all possible supports and grants accessible.

It is my pleasure to thank the secretary of Doctorate School Sig. Andrea Zanzani who made all the documentation work easier and helped me, for my permit of stay and other regulations and norms.

It is my pleasure to pay tribute to *Dr. Gea Donzelli*, who has been my Italian Teacher. My heartfelt thanks to *Dr. Fabio Giavazzi*, my friend, and more than that he is my brother who has shown keen interest on my welfare. It is my pleasure to thank *Carlotta Munforti*, *Mariachaira Rossetti*, *Federica Ermetici*, *Michele Bernardin*, *Tiziano Sanvito*, *Michele Manfredda*, *Elisa Tomborini*, who made my days happier in Milan. Though I had shared few hours with you people, but those moments are unforgettable in my life.

I thank my colleagues *Berihu Teklu*, *Dr. Cheedom Ozkan*, for being supportive during my course.

I thank my Indian friends *Benjamin Tamilselvan Nachimuthu*, *Dr. Dinesh Velayutham*, *Dr. Maharaja Ponnaiah*, *Dr. Pauline Sandra*, *Dr. Mlind Dangate*, *Rama Dangate*, *Adkar Purshothama Charith*, *Mathew*, *Siva* for being supportive in Milan.

Where would I be without my family? My parents deserve special mention for their inseparable support and prayers. My father *Mr. K. P. Velu*, who has been the role model of my life, taught and shown me the joy of intellectual pursuit ever since I was a child. My mother *Mrs. V. Devi*, has sincerely raised me by her love, care and affection. My beloved brother *Mr. K. P. V. Sathish* and brother-in-law *Er. T. Muralikrishnan* for being supportive.

Words fail to express my appreciation to my fiance *T. Kalaivani* without whom I would not achieved these heights in my life. I owe her for being unselfishly let her love, care, passions and ambitions collide with mine. I would also thank my uncle *Mr. V. R. Thangavel* and my aunt *Mrs. T. Muthulakshmi* for letting me and accepting me as a member of their family.

Finally, I would, like to *thank everybody* who was important to the successful realization of thesis, as well as expressing my apology that I could not mention personally one by one.

*-Sabareesh Velu*

# CONTENTS

# Chapter 1

# Introduction

# Chapter 1

# Introduction

---

Scattering methods are key techniques for studying the soft condensed matter systems such as colloids, polymers, surfactants, gels and biological macromolecules [1,2]. Traditionally, the scattering techniques measure the scattered intensity as a function of angle. Anyway, since the process involves a scattered wave, one could ask if any additional information can be gained by measuring the actual phase delay between the incoming plane wave and the scattered wave.

Although it is not widely recognized, the Mie theory does predict a phase difference between the incoming plane wave and the scattered wave at zero angle. Most of that phase change occurs when the scatterer optical thickness approximately varies between a small fraction of the wavelength up to the wavelength itself, the total change becoming asymptotically equal to $\pi/2$ for larger thickness [3]. So by measuring phase delay in principle one could measure the optical thickness of the scatterers, a quantity of interest in phase transition of colloidal studies and not easily obtainable with any classical scattering methods.

Particularly interesting is the case of small absorbing scatterers, as absorption rather than scattering is the prevailing mechanism, and then the phase change occurs even at smaller diameters thus opening the possibility of studying the particles well in the deep sub – micron regime. In spite of these potentially attractive perspectives, no experimental method based on measuring the phase delay of the scattered wave has been developed so far. The reason for this disappointing situation is not accidental and will be detailed further.

The zero angle scattered wave plays an essential role in the celebrated Optical Theorem that relates the total extinction cross section (including both scattering and absorption) just to the complex amplitude of the scattered wave. In an authoritative paper [4] where the possibilities of an experimental verification of Optical Theorem were discussed, it was pointed out that a proof could hardly come from an optics experiment.

One must agree to this statement if one receives the scattering intensity from an assembly of particles as a Bragg reflection from a three dimensional random grating. Accordingly, scattered waves at any angle then behave as a circular Gaussian process, and phases are equally distributed in the interval ($-\pi$, $+\pi$) [5], therefore giving no cue on the behaviour of the wave scattered at exactly zero angle. So a proof of the Optical Theorem would require the feat of telling apart an incoming wave from its weak replica (exact down to the finer details), and gauge the phase in between.

In this thesis, Near Field scattering method is exploited just for the determination of this elusive phase of the scattered wave. Having determined the phase of the scattered wave, the experimental verification of Optical Theorem can also be attempted.

The Near Field scattering method also allows to unambiguously identify the fraction of the total scattered power that is due to single scattering alone, thus allowing the application of the method even when multiple scattering is rampant (beam attenuation up to 99%) has been used here.

In addition, the method can generate dynamic scattering data as well; it provides an elegant and efficient way to filter out stray multiple scattering artefacts. Other interesting practical applications will also be discussed.

# Chapter 2

# Experimental Setup

# Chapter 2

# Experimental Setup

The pictorial representation of the technique is shown in figure 2.1. The experimental setup consists of two subsystems such as cell and optical setup. The two subsystems are detailed separately in the succeeding paragraphs.

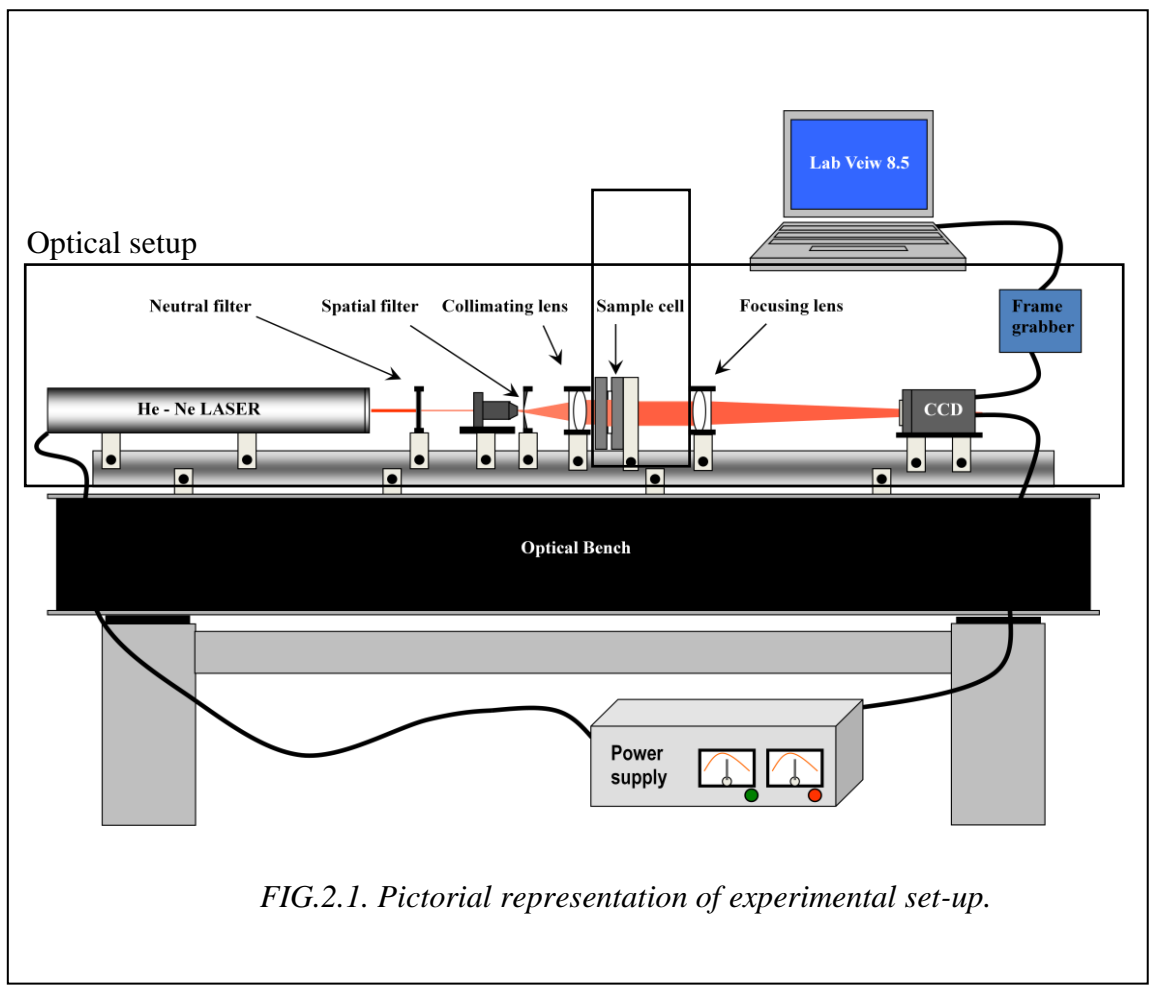

*FIG.2.1. Pictorial representation of experimental set-up.*

## 2.1 Cell

A key role in the experimental set-up is played by the confinement cell. The task of the cell is to confine a thin vertical layer of the sample. For our experiments, the cell must provide a secure confinement of the sample, avoiding bubble formation, chemical attack of the surfaces and leaking of the sample. Also, the cell must provide optical access through the sample keeping the illuminated area as large as possible. Keeping all this in mind, the cell is custom-made.

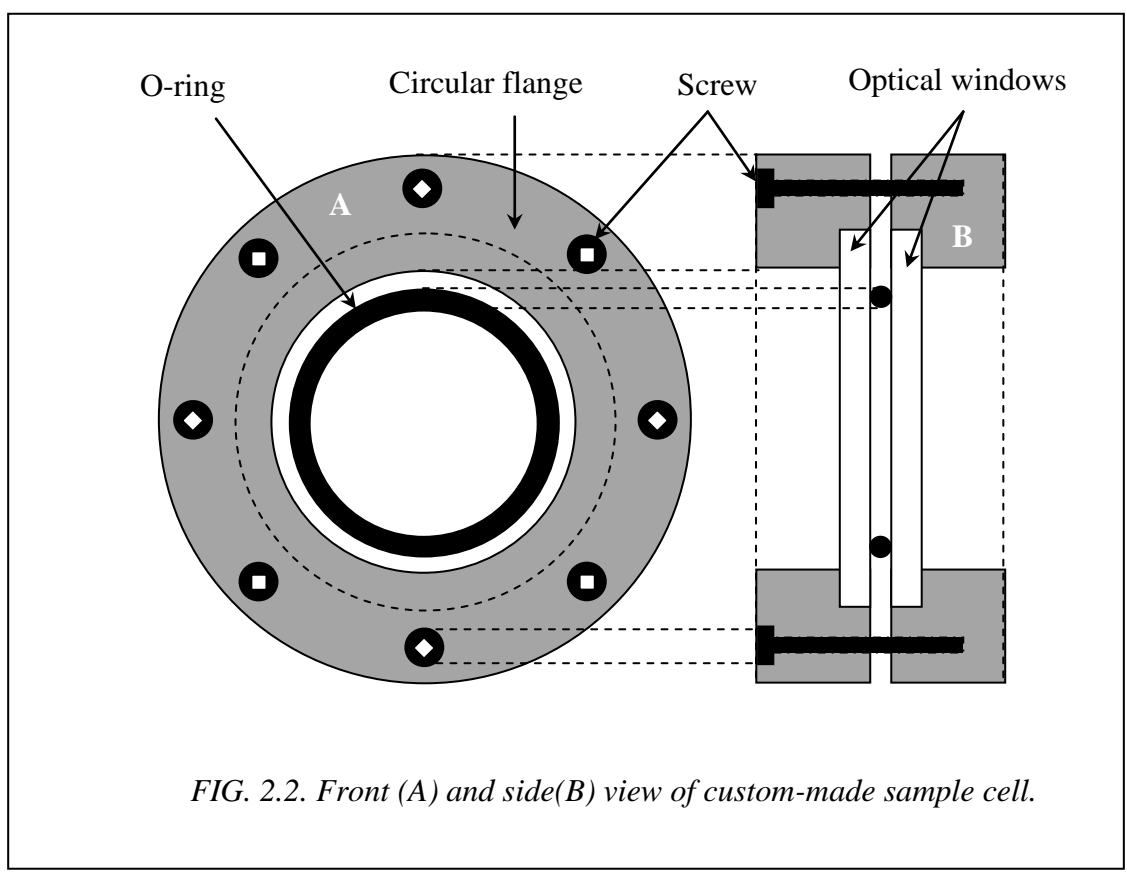

*FIG. 2.2. Front (A) and side(B) view of custom-made sample cell.*

Two circular optical windows of diameter 5cm and thickness 0.4cm are hosted within circular flanges (made up of plastic material called Ertacetal C. The optical windows are of optical quality. Leak tight confinement of the sample is provided by an O-ring (diameter 30 mm and thickness 0.3cm and when compressed 2.7 mm) inserted between the optical windows as shown in figure 2.2. The gasket is compressed at the prescribed value regulated by the presence of plastic spacers. We used ordinary nitrile O-rings since the sample to be confined is non aggressive aqueous solution.

The thickness of plastic spacers is chosen in accordance with the prescribed values for the compression of the O-ring. The compression of the spacer is provided by 8 screws distributed along the circumference of the flange. Precautions should be taken to apply torque homogenously to all the screws; otherwise this will result in a wedging of the sample layer. To avoid this following procedure should be adopted while mounting the cell. A torque wrench is used to tighten the screws. After all the screws are tighten, the procedure is repeated by increasing the torque until the plastic spacers are compressed. Finally the cell is assembled as described and ready to load the sample to be analysed into the assembled cell.

Two syringe needles are fixed opposite to each other within O-ring fixed in between the optical windows. Then the required amount of sample is taken in syringe and connected to one of the needle. To the other needle, an empty syringe is connected without any air within it. Apply gentle pressure to the syringe containing the sample such that the sample will flow into the mounted cell. Precautions should be taken to fill the cell without air bubbles within the sample. Once the sample is loaded within the cell without any bubbles, the two needles are gently removed.

## 2.2 Optical setup

The optical set-up consist of laser source, spatial filter, collimating lens, sample cell *C*, collection lens and 2-D charge coupled device (CCD) sensor *S*.

## Laser source:

The low angle scattering technique works in heterodyne scheme and therefore one should take into account that at sufficiently large scattering angles, the phase relation between the interfering beams will be lost (loss of temporal coherence) or can be strongly affected by the finite size of the source (loss of spatial coherence). Hence we tempted to use a source with good coherence properties like a laser.

The laser source used in our setup is the cylindrical helium – neon (He-Ne) laser from Melles Griot. The output power of the laser is 30 mW maximum at wavelength $\lambda = 632.8$ nm and linearly polarised. The beam diameter ($1/e^2$) of the output laser is 0.65 mm.

## Spatial filter:

The internal mirror of He-Ne laser is designed to operate in $TEM_{00}$ mode and those that are multimode and cannot change to $TEM_{00}$ mode. The spatial filter consists of a microscope objective, a pinhole aperture and a positioning mechanism. Their combinations are used to select single transverse mode generally used in scattering experiments.

## Collimating optics:

The collimating optics consists of one doublet lens $f_1$ placed after the spatial filter to provide a collimated beam that impinges on the sample. The choice of lens focal length and distance between the spatial filter and lens depends on the requirements of laser beam diameter. In our optical set-up we used the doublet lens of focal length $f_1 = 30$ cm such that the collimated laser beam diameter, $D \sim 10$ mm ($1/e$ full width). The laser beam diameter is determined using knife - edge method [1].

Collection optics:

The collection optics includes a doublet lens $f_2$ and CCD sensor. Two different schemes are possible called Extra Focal and Intra Focal based on the position of the visualization plane with respect to the focal plane of the collecting lens. In extra focal scheme the sensor is placed beyond the focal length of the lens while in intra focal scheme the sensor is placed within the focal length of the lens.

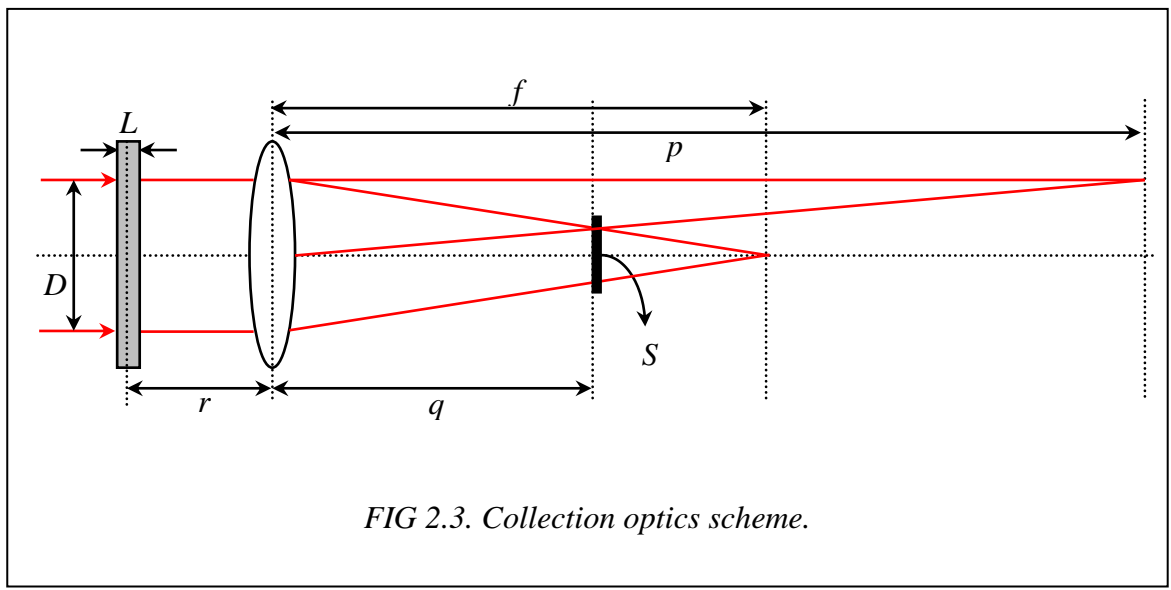

*FIG 2.3. Collection optics scheme.*

In our set-up, intra focal scheme is used taking advantage of reaching ultra low angles in less space otherwise using extra focal scheme requires huge space to reach such ultra small angles. Figure 2.3 shows the optical set-up of our collection optics. The doublet lens is placed at a distance *r* from the sample cell and the sensor is placed at a distance *r+q* from the sample cell where the *q* is the lens – sensor distance.

The distance *p* and *q* can be related as

$$\frac{1}{p}+\frac{1}{q}=\frac{1}{f} \qquad \text{eq. (2.1)}$$

Then the effective distance between the sample and sensor, $z = p+r$ with a magnification factor $M = q/p$ and alternately, the sample - sensor effective distance and the magnification factor can be related as

$$M = \frac{f-q}{f} \qquad \text{eq. (2.2)}$$

$$z = \frac{q}{M} + r \qquad \text{eq. (2.3)}$$

The focal length of the collecting lens is $f = 50$ cm, the sample - collecting lens distance is $r = 16.8$ cm, collecting lens - sensor distance is $q = 31.5$ cm and thus the magnification factor is $M = 0.37$ and the effective sample-sensor distance is $z = 102$ cm.

Notice, the distance $q$ is shorter than the focal length $f$ of lens and hence the speckles reproduced on the sensor plane are virtual and the distance $z$ turns out to negative.

## Sensor:

The sensor used is 8-bit Jai CV-M300 industrial monochrome CCD sensor. The CCD sensor is interfaced to the computer by means of frame grabber (National instrument, model PCI-1422) and controlled using LabVIEW (National Instruments) software.

The sensor resolution is 752 x 582, and the dimension of the single pixel is 11.6 µm. Only a square portion of 512 x 512 pixels is used for data analysis and thus the area of the sensor is 0.59 x 0.59 cm$^2$. Due to the magnification factor, the sensor records the speckle images $I(x, y)$ corresponding to an area of 1.6 x 1.6 cm$^2$ of sample cell.

The 2-D Fourier transform is performed on the recorded speckle image $I(x, y)$ to a transformed image the Fourier space $I(q_x, q_y) = F [I (x, y)]$ in which the capital letter represents the spatial Fourier transform (F), $q_x$, $q_y$ being the Fourier vectors associated with the spatial frequencies $I_x$ and $I_y$ ($q_x = 2\pi I_x$, $q_y = 2\pi I_y$). The maximum and minimum detectable wave vectors given by

$$q_{\min} = \frac{2\pi}{n} M \qquad \text{eq. (2.4)}$$

$$q_{\max} = \frac{n}{2} q_{\min} \qquad \text{eq. (2.5)}$$

where $n$ is number of pixels. Using eq. 2.4 and eq. 2.5, the theoretical minimum detectable wave vector is $q_{min}$ = 3.9 cm$^{-1}$ and the maximum detectable wave vector is $q_{max}$ = 998.4 cm$^{-1}$. The theoretical range is experimentally limited ($q_{min}$ = 23 cm$^{-1}$ and $q_{max}$ = 500 cm$^{-1}$) by noise.

## 2.3 Working Principle of the Technique

The technique operates in the deep Fresnel region ($z \ll D^2/\lambda$; where $z$ is the sample - sensor distance, $D$ is the beam diameter and $\lambda$ is the wavelength of the light) such that the near field scattering conditions are met [6, 7].

It also operates in heterodyne detection scheme, i.e., the transmitted light is superimposed with the much weaker scattered light and the resulting interference pattern is recorded using a 2-D digital camera as speckle images.

The technique works for low angles ($\theta < \theta^* = (\lambda/L)^{1/2}$ being $L$ is the sample thickness) such that the impinging light on the scattering medium diffracts light equally into positive and negative orders that are in phase. This is referred to as Raman-Nath scattering regime [8]. For higher angles ($\theta > \theta^*$) the Bragg condition holds, where the lights diffracted into positive and negative orders are not in phase and the scattering has 3-D character [9]. In the Raman-Nath regime the scattering is from a 2-D phase gratings and any scattering event always generates two waves at symmetric angles [9]. The three wave interference onto the sensor has a low contrast speckle distribution given by the superposition of many interference patterns between the transmitted beam and the spherical waves scattered by each particle.

Let a particle of arbitrary shape and composition being in the origin be illuminated by a plane wave written as

$$A_O = \exp(-ikz + i\omega t) \qquad \text{eq. (2.6)}$$

where $\lambda$ is the radiation wavelength, $k = 2\pi/\lambda$, $A_o$ is the amplitude of the incident wave.

The complex amplitude of the scattered wave at a distance $r$ can be written as

$$A = S(\theta)\frac{\exp(-ikz + i\omega t)}{ikr} \qquad \text{eq. (2.7)}$$

By combining eq. (2.6) and (2.7) can be written as

$$A(r,\theta) = S(\theta)\frac{\exp(-ikz + ikz)}{ikr} A_O \qquad \text{eq. (2.8)}$$

where ($r$, $\theta$) is the detector position, $S(\theta)$ the scattering amplitude. The intensity on the sensor is due to the three wave interference pattern is given by

$$I(x,y) = |A_O|^2 \frac{2A_O|S(0)|}{kz}\cos\left[k\frac{x^2 + y^2}{2z} - \phi + \frac{\pi}{2}\right] \qquad \text{eq. (2.9)}$$

The eq. (2.9) is evaluated at a distance $z$, and the phase delay $\phi = Arg[S(0)]$ has been introduced. Although the intensity distribution generated by a collection of many scatterers appears as a completely stochastic signal, the fact that it is made by circular interference fringes becomes evident when the spatial power spectrum of the intensity distribution is taken as

$$S(q_x, q_y) = \frac{4\pi^2}{k^2} |S(0)|^2 \sin\left[\frac{q^2 z}{2k} - \phi\right] \qquad \text{eq. (2.10)}$$

where $q^2 = q_x^2 + q_y^2$. Both functions eq. (2.9) and (2.10) have an infinite sequence of oscillations, and though not widely appreciated, they possess a unique property, namely that all the power under the $2i$-th fringe of $S(q_x, q_y)$ comes soley from the $i$-th fringe of $I(x, y)$. So, if a given fringe in the interference pattern shifts at smaller diameters then by increasing the distance $z$ the same occurs to the oscillation of the same order in the power spectrum. Incidentally, this also explains why $\phi$ subtracts up in the argument in both the trigonometric expressions of $S(q_x, q_y)$ and $I(x, y)$. As each fringe is associated to a $2\pi$ phase change between the scattered spherical wave and the incoming wave, the phase delay $\phi$ at zero angle can be very accurately extrapolated by fitting the entire sequence of oscillations.

## 2.4 Advantages of the Technique

The advantages of the techniques are:

- immune to the stray lights, can access wave vectors as small as few $cm^{-1}$ which corresponds to milli-degrees in scattering angles
- the optical set-up of the technique is simple and do not require accurate alignment

- the technique use CCD detection and provides simultaneous measurements for many different wave vectors and thus good statistics can be obtained in short times.

*Chapter 3*

*Statistical Image Analysis*

# Chapter 3

# Statistical Image Analysis

A set of *N* frames (speckle images) with delay time $\tau$ are acquired. The data acquisition is controlled by the computer which is interfaced with our sensor using LabVIEW program. Then the acquired speckle images are statistically analysed to recover the phase of the scattered waves from the power spectrum of our samples.

For the recorded speckle images, the double frame statistical analysis is used to obtain the static and dynamic power spectrums. In the succeeding paragraphs the double frame statistical analysis [7] is discussed in detail.

## 3.1 Differential double frame analysis:

The intensity distribution of the recorded speckle images is composed of strong static electric field $e_O(r)$ associated with the transmitted beam plus the stray light acts as a local oscillator and the much weaker time dependent scattered field $e_S(r)$. We show a speckle image i.e., intensity distribution $I(r,t)$.

The speckle image is recorded with a magnification $M = 0.37$. The magnification factor *M* is less than unity due to focused setup. The resulting intensity distribution $f(r,t)$ can be written as

$$I(r,t) = I_O(r) + e_O(r)e_S^*(r,t) + e_O^*(r)e_S(r,t) + I_S(r,t) \qquad \text{eq. (3.1)}$$

$$I(r,t) = I_O(r) + 2\text{Re}\{e_o^*(r)e_S(r,t)\} + I_S(r,t) \qquad \text{eq. (3.2)}$$

where *r* is the two-dimensional vector (*x*, *y*). The contribution due to the interference between the scattered waves (homodyne term) is negligible ($|e_S| \ll |e_O|$) and the term $I_S(r,t) = |e_S(r,t)|^2$ can be dropped in eq. (3.1) & eq. (3.2). The optical background term $I_O(r)$ depends on *r* and accounts for the spatial dependence of the spurious stray light which is present in any low angle light scattering system. The terms $e_S(r,t)$ and $e_S^*(r,t)$ vary with time and are zero averaged.

Suppose a set of frames acquired corresponding many independent sample configurations and let $\tau$ be the temporal distance between each frame. If $I_1(r,t)$ and $I_2(r,t)$ are the taken at time *t* and *t*+$\tau$ respectively, the intensity distribution can be given as

$$I_1(r,t) = I_O(r) + 2\text{Re}\{e_o^*(r)e_1(r,t)\} \qquad \text{eq. (3.3)}$$

$$I_2(r,t,\tau) = I_O(r) + 2\text{Re}\{e_o^*(r)e_2(r,t+\tau)\} \qquad \text{eq. (3.4)}$$

The differential heterodyne signal can be extracted by taking the difference of eq. 3.3 and eq. 3.4 and given as

$$\delta I(r,t,\tau) = I_2(r,t+\tau) - I_1(r,t) \qquad \text{eq. (3.5)}$$

Therefore eq. (3.5) can be written as

$$\delta I(r,t,\tau) = 2\text{Re}\{e_o^*(r)e_2(r,t+\tau) - e_o^*(r)e_1(r,t)\} \qquad \text{eq. (3.6)}$$

The eq. (3.6) shows that the difference of two images is independent of $I_o(r)$ and the difference image containing only the signal. To correct the Gaussian nature of strong electric field, the difference image is divided by $I_2(r, t+\tau) + I_1(r,t)$ and eq. (3.6) modifies as

$$\delta I(r,t,\tau) = \frac{2}{I_o} \text{Re}\{e_o^*(r) e_2(r,t+\tau) - e_o^*(r) e_1(r,t)\} \qquad \text{eq. (3.7)}$$

Then the result is multiplied by the average intensity of the speckle image to maintain the actual intensity level of the recorded image in the analysis.

$$\delta I(r,t,\tau) = \langle I_o \rangle \frac{2}{I_o} \text{Re}\{e_o^*(r) e_2(r,t+\tau) - e_o^*(r) e_1(r,t)\} \qquad \text{eq. (3.8)}$$

$$\delta I(r,t,\tau) = 2 \text{Re}\{e_o^*(r) e_2(r,t+\tau) - e_o^*(r) e_1(r,t)\} \qquad \text{eq. (3.9)}$$

The ratio image $[I_2(r,t,\tau) - I_1(r,t) / I_2(r,t,\tau) + I_1(r,t)]$ containing the signal is computed for its Fourier components then taking Fourier transform ($I$ is replaced as $S$) of eq. (3.9) can be written as

$$\delta S(r,t,\tau) = 2 \text{Re}\{E_o^*(-q) * E_2(q,t+\tau) - E_o^*(-q) * E_1(q,t)\} \qquad \text{eq. (3.10)}$$

where $q \equiv (q_x, q_y)$, $q_x$ and $q_y$ being the Fourier vectors associated with the spatial frequencies $I_x$ and $I_y$ ($q_x = 2\pi I_x$, $q_y = 2\pi I_y$).

The eq. (3.10) shows that the $q$ component of the signal observed on the sensor plane is the result of two contributions deriving from the interference between the reference beam and the three-dimensional plane waves scattered with two vectors $k_S^+ = (q, k_z)$ and $k_S^- = (-q, k_z)$ and amplitudes $E_s(q,t)$ and $E_s(-q,t)$, respectively.

The two-dimensional power spectrum of the heterodyne signal is obtained by squaring eq. (3.10), given as

$$|\delta S(q,t,\Delta t)|^2 = |\alpha_2|^2 + |\alpha_1|^2 - \alpha_2 \alpha_1^* - \alpha_2^* \alpha_1 \qquad \text{eq. (3.11)}$$

where

$$\alpha_2 = 2\text{Re}\{E_o^*(-q) * E_2(q,t+\tau)\}$$
$$\alpha_1 = 2\text{Re}\{E_o^*(-q) * E_1(q,t)\} \qquad \text{eq. (3.12)}$$

Solving eq. (3.11) we can write eq. (3.12) as

$$|\delta S(q,t,\tau)|^2 =$$
$$\left|E_o(q) * E_2^*(-q,t+\tau)\right|^2 + \left|E_0^*(-q) * E_2(q,t+\tau)\right|^2 +$$
$$\left|E_o(q) * E_1^*(-q,t)\right|^2 + \left|E_0^*(-q) * E_1(q,t)\right|^2 -$$
$$2\text{Re}\left\{E_o(q) * E_2^*(-q,t+\tau) \cdot E_o(q) * E_1^*(-q,t) + E_o^*(-q) * E_2(q,t+\tau) \cdot E_0^*(-q) * E_1(q,t)\right\} +$$
$$2\text{Re}\left\{E_o(q) * E_2^*(-q,t+\tau) \cdot E_0^*(-q) * E_2(q,t+\tau) + E_o(q) * E_1^*(-q,t) \cdot E_0^*(-q) * E_1(q,t)\right\} -$$
$$2\text{Re}\left\{E_o(q) * E_2^*(-q,t+\tau) \cdot E_0^*(-q) * E_1(q,t) + E_o^*(-q) * E_2(q,t+\tau) \cdot E_o(q) * E_1^*(-q,t)\right\}$$

eq. (3.13)

In eq. (3.13), the terms appearing in lines 1, 2 and 3 involve the products of the fields scattered with the same wave vector (either $+q$ or $-q$). The lines 1 and 2 can be directly related to mean square amplitude function of scattering particles or the scattered intensity distribution $|S(\theta = 0)|$. The line 3 directly relates the field-field correlation function $G_E(q,\tau)$.

Lines 4 and 5 involve the products of the fields scattered with opposite wave vectors ($+q$ and $-q$) called phase correlated terms responsible for the deep oscillations.

For isotropic samples we can take the azimuthal average of eq. (3.13) to obtain the radial power spectrum

$$S(q_x, q_y) = \left\langle |\delta S(q,t,\tau)|^2 \right\rangle_{t,q} \qquad \text{eq. (3.14)}$$

In which the $t$ is average of large number of independent sample configurations and $q$ average is carried out over all the vectors such that $q = \sqrt{q_x^2 + q_y^2}$. The eq. (3.14) can be re-written as

$$S(q_x, q_y) = A\left\{|S(0)|^2 - G_E(q,\tau)\right\} \cdot \sin^2\left(\frac{q^2 z}{2k} - \phi\right) + B(q) \qquad \text{eq. (3.15)}$$

where $A$ is constant, $\phi$ is the phase lag of scattered wave with respect to the incident wave and $B(q)$ is a measurable noise background, due to almost entirely to electronic and shot noise in the camera.

If the delay time $\tau$ is large enough than the correlation time $\tau_c$, then $G_E(q,\tau) \rightarrow 0$, thus

$$S(q_x, q_y) = A \cdot |S(0)|^2 \cdot \sin^2\left(\frac{q^2 z}{2k} - \phi\right) + B(q) \qquad \text{eq. (3.16)}$$

The eq. (3.16) have an infinite sequence of oscillations, As each oscillation is associated to a $2\pi$ phase change between the scattered spherical wave and the incoming wave, the phase delay $\phi$ at zero angle can be very accurately extrapolated by fitting the entire sequence of oscillations.

*Chapter 4*

*Near Field Scattering: Experimental Results*

# Chapter 4

# Near Field Scattering: Experimental Results

---

In this chapter, the experimental results of extreme low angle scattering measurements performed on clouds of many mono disperse particles are presented. The experimental method exploits the Near Field scattering scheme [6, 7].

## 4.1. Technique

The basic scheme of Near Field scattering technique is shown in figure. 4.1. A spatially filtered and large collimated Gaussian laser beam, $D \sim 10$ mm in diameter (1/e full width), passes through a plane windows cell (4 cm in diameter, 2.7 mm in thickness, windows being of optical quality) containing the sample. A fraction of light beam is scattered by the sample and superimposes with the transmitted beam in the forward direction.

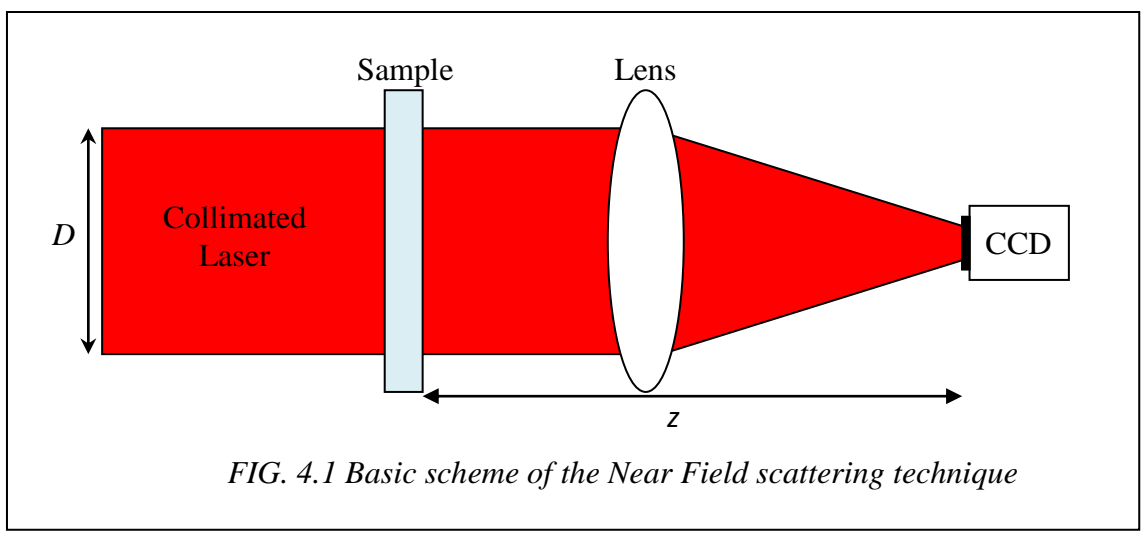

*FIG. 4.1 Basic scheme of the Near Field scattering technique*

The receiving optics is a doublet lens 50 cm focal length and 5 cm in diameter. A two dimensional (2-D) CCD sensor 512 × 512 with pixel size $l_{pixel}$ = 11.4 μm is placed 31.5 cm away from the lens, before the beam is actually brought to a focus. The receiving optics essentially folds the optical path, and reproduces what a sensor would see if placed 102 cm away from the cell, without receiving lens and the area covered by the sensor corresponds to an area of 1.6 × 1.6 $cm^2$ inside the cell (Chapter 2).

The resulting intensity on the sensor has a low contrast speckle distribution due to the superposition of many interference patterns between the intense transmitted beam and the faint spherical waves scattered by each particle. The speckle distribution is recorded and statistically analysed according to the Near Field scattering data reduction schemes [7] to recover the scattered intensity distribution.

The experimental set up and description of the technique is detailed in Chapter 2. The statistical analysis is based on differential double frame approach and the procedure is explained in Chapter 3.

## 4.2. Experimental Results

### 4.2.1. Experimental procedure and data reduction scheme:

The measurements are carried out on aqueous solution of polystyrene colloidal particles of diameter 2 μm (Duke scientific corporation). A set of $N$ images of near field speckle distributions generated by the sample are recorded with a delay time $\tau$ = 200 ms. The recorded heterodyne signal (figure 4.2) is composed of strong static signal electric field $e_O(r)$ due to the main beam intensity on which a weak time dependent scattered field $e_S(r,t)$ is superimposed can be written as

$$I(r,t) = i_O(r) + e_O(r)e_S^*(r,t) + e_O^*(r)e_S(r,t) + i_S(r,t) \qquad \text{eq. (4.1)}$$

where $r = (x, y)$ is the two-dimensional vector and the apex * indicates the complex conjugate. The contribution due to the interference between the scattered waves (homodyne term) is negligible ($|e_S| < |e_O|$) and dropped in eq. (4.1).

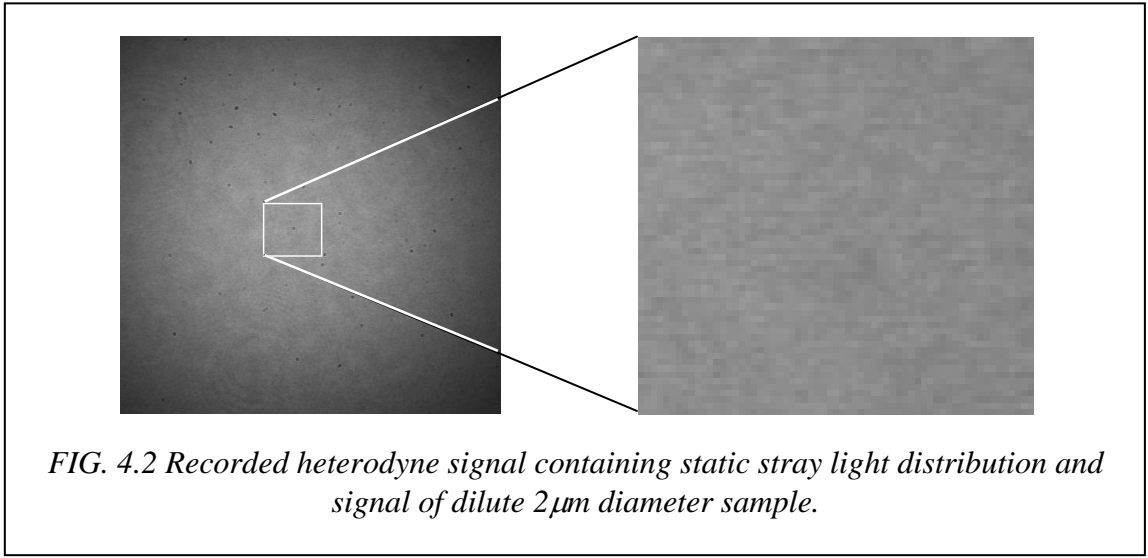

*FIG. 4.2 Recorded heterodyne signal containing static stray light distribution and signal of dilute 2μm diameter sample.*

In the data reduction, time differences between the recorded images are generated, the time delay being varied between 0.2 sec up to 500 sec. The advantage of taking time differences of images allows removing any static stray light contribution. Rationing to the sum then equalizes the average of the heterodyne signal. The averaged heterodyne signal is then Fourier transformed and squared to recover the 2-D power spectrum and azimuthally averaged power spectrum.

The Near Field speckles fluctuates randomly due to the motion of colloidal particles. If the time differences of two images are large enough, then the speckles are completely de-correlated and the obtained power spectrum gives the scattered intensity distribution (static structure). The speckles are de-correlated much faster as compared to the time constant of its diffusive motion. The motion of colloidal particles inside the cell is due to convective motions.

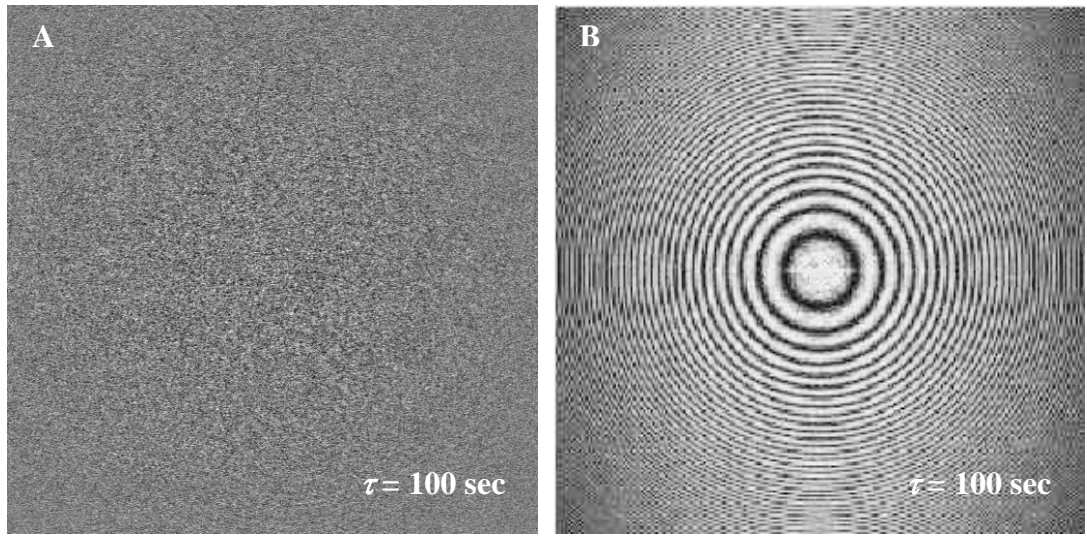

*FIG. 4.3 (A) De-correlated differential image and (B) corresponding time averaged 2-D power spectrum of dilute 2μm sample.*

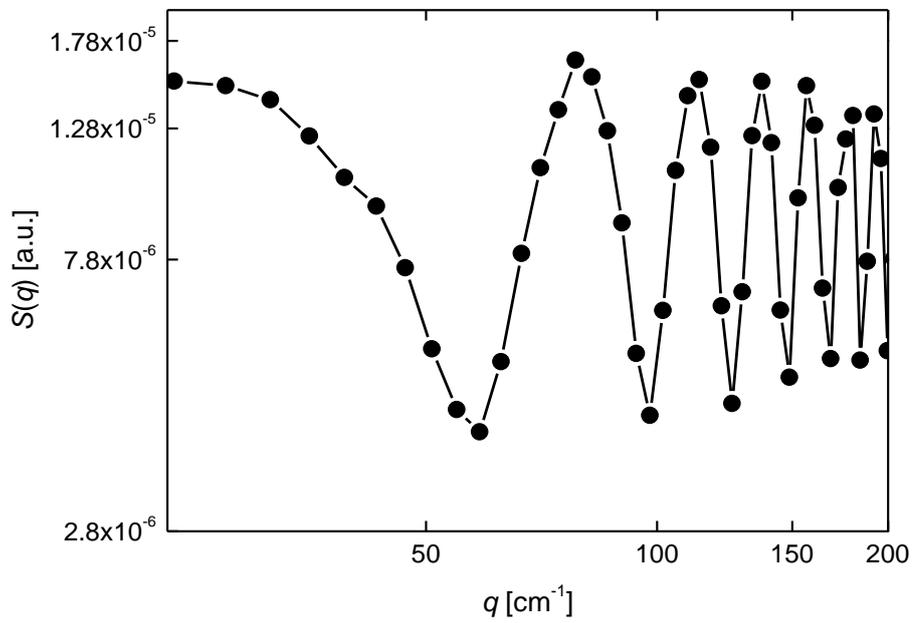

*FIG. 4.4 Azimuthally averaged power spectrum of dilute 2μm sample.*

The figure 4.3(A) shows the de-correlated differential heterodyne speckle image and figure 4.3(B) shows the time averaged 2-D power spectrum of de-correlated speckle images of dilute 2 μm sample with a volume fraction $\Phi = 10^{-5}$.

The scattered intensity for 2 μm diameter polystyrene particles in water as a function of scattering wave vector in the range of interest (20 – 200 cm$^{-1}$) is essentially flat as calculated with Mie theory. The recovered time averaged 2-D power spectrum contains a sequence of circular structures that exhibit a peculiar oscillatory modulation in the azimuthally averaged power spectrum (see figure 4.4).

This is because the Near Field scattering technique works for so small angles such that the Raman-Nath scattering conditions [8] are full filled (Chapter 2). In this regime, the scattering is due to 2-D phase gratings and any scattering event always diffracts light equally into positive and negative orders that are in phase. The phase locked scattered waves at symmetric angles and it is the three wave interference that generates this peculiar oscillatory modulation.

If $A_o$ is the amplitude of the incident wave, $S(\theta)$ is the scattering amplitude function of the scattering particle as a function of the azimuthal angle $\theta$, $\lambda$ the radiation wavelength, $k = 2\pi/\lambda$. Thus the corresponding interference pattern with the transmitted beam on the sensor at a given time $t$ is given by

$$I(x,y) = |A_o|^2 + \frac{2A_o |S(0)|}{kz} \cos\left[ k\frac{x^2 + y^2}{2z} - \phi + \frac{\pi}{2} \right] \qquad \text{eq. (4.2)}$$

Here the interference pattern is evaluated at a distance $z$ along in the forward direction, such that the paraxial approximation holds and $S(\theta) \sim S(0)$, and the phase delay $\phi = \text{Arg}[S(0)]$ is introduced, determining the fractional order at the centre of the fringe pattern. Although the intensity distribution generated by a collection of many scatterers appears as a completely stochastic signal, the fact that they are made up by circular interference fringes becomes evident when the spatial power spectrum $S(q_x, q_y)$ of the intensity distribution $I(x, y)$ is taken. Indeed, the power spectrum does not contain just the information about the position of the centres of the fringe systems, thus being identical to the power spectral density of one fringe system:

$$S(q_x, q_y) = \frac{4\pi^2}{k}|S(0)|^2 \sin^2\left[\frac{q^2 z}{2k} - \phi\right] \qquad \text{eq. (4.3)}$$

where $q^2 = q_x^2 + q_y^2$. Both the functions eq. (4.2) and eq. (4.3) have an infinite sequence of oscillations, and they possess a unique property, namely that all the power under the *2i-th* oscillation of $S(q_x, q_y)$ comes from the *i-th* fringe of $I(x, y)$ alone.

### 4.2.2. Measurement on colloidal particles:

The measurements are carried out on mono disperse polystyrene spheres with diameters 0.2, 0.6, 1, 3, and 5 µm in aqueous suspension (Duke scientific corporation).

The volume fractions is $\Phi = 5 \times 10^{-5}$ for 0.6, 1, 3 and 5 µm samples. For 0.2 µm the volume fraction is $\Phi = 5 \times 10^{-4}$. All the measurements are performed at room temperature. The de-correlated differential heterodyne speckle image ($\tau = 100$ sec) for various diameters of colloidal particles and corresponding time averaged 2-D power spectrums are shown in figure 4.5 (A) and (B) respectively. The azimuthally averaged power spectra are shown in figure 4.6. The graphs are plotted in the same relative scale.

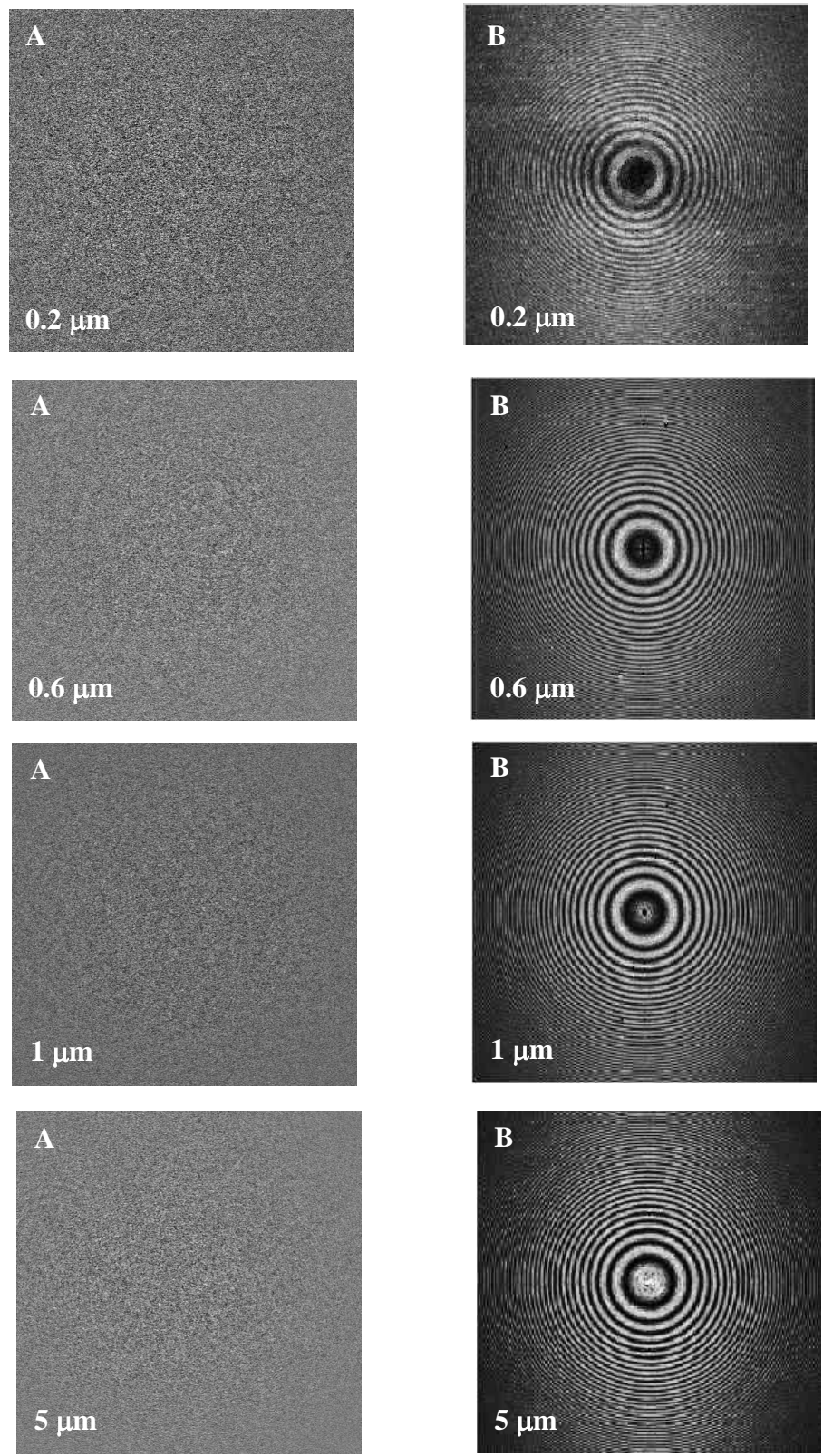

*FIG. 4.5(A) De-correlated differential image for 0.2, 0.6, 1 & 5 µm samples and (B) corresponding time averaged 2-D power spectrum.*

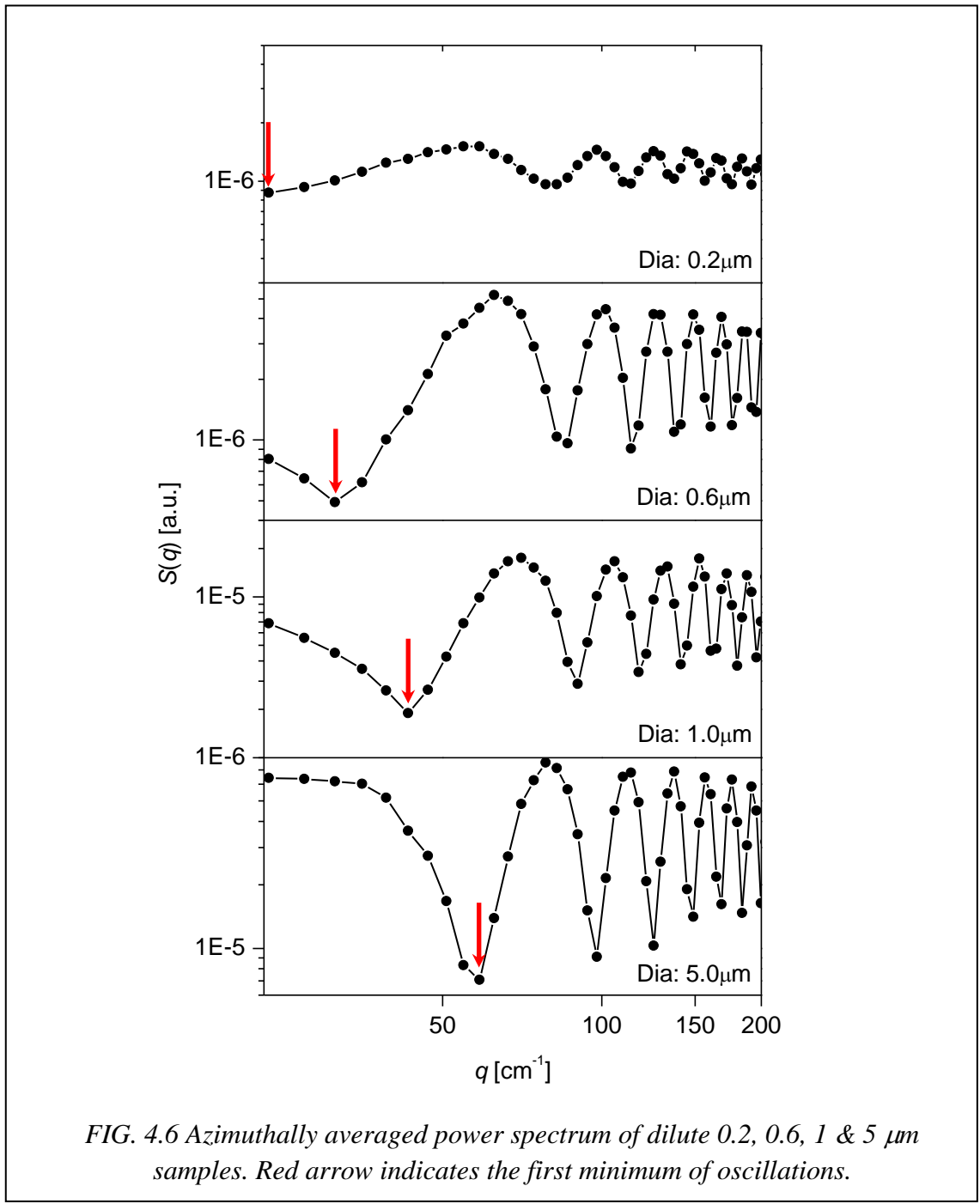

*FIG. 4.6 Azimuthally averaged power spectrum of dilute 0.2, 0.6, 1 & 5 µm samples. Red arrow indicates the first minimum of oscillations.*

The recovered time averaged 2-D power spectrum contains sequence of circular structures that exhibits the deep oscillations in the azimuthally averaged power spectrum. Notice that the maxima and minima in the azimuthally averaged power spectrum recovered for various diameters of sample shift to higher wave vectors for larger diameters (see the red arrow indicating the first minimum).

As each fringe is associated with a $2\pi$ phase change between the scattered spherical wave and the incoming wave, the phase delay $\phi$ at zero scattering angle can be very accurately extrapolated by fitting the entire sequence of oscillations at larger angles in the power spectral density obtained with many identical particles. The figure 4.7 shows the fitting of eq. (4.3) for power spectrum of 2 μm ($\Phi = 5 \times 10^{-5}$) and directly recovers the phase delay.

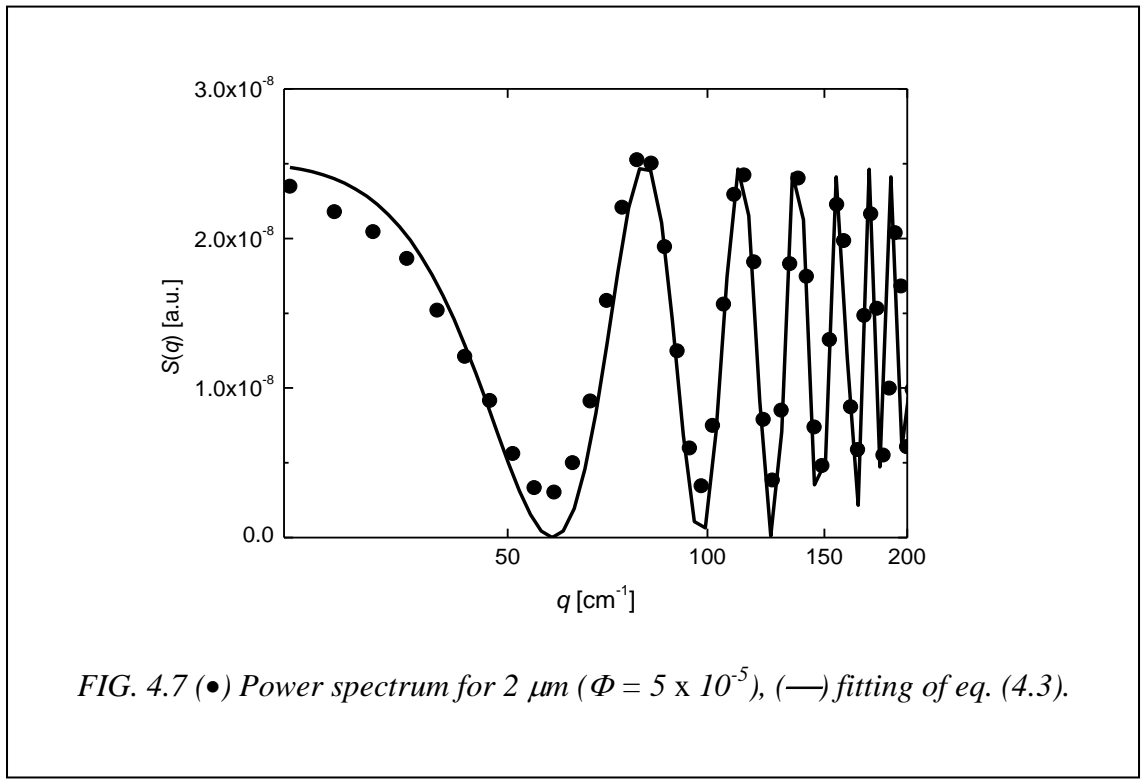

*FIG. 4.7 (●) Power spectrum for 2 μm ($\Phi = 5 \times 10^{-5}$), (——) fitting of eq. (4.3).*

Similarly, the phase delays are measured for various diameters of samples and listed in table 4.1. The figure 4.8 shows the measured phase lag as a function of particle diameter. The oscillation in the recovered power spectrums decreases at higher $q$ values (see figure 4.6). This is because of limited spatial resolution of the sensor but this is of little relevance here as pure phase information is sought.

| Sample | Measured phase delay ($\phi$) | Error |
|---|---|---|
| 0.2 μm | 0.14 | 0.03 |
| 0.6 μm | 0.519 | 0.02 |
| 1.0 μm | 0.95 | 0.017 |
| 2.0 μm | 1.753 | 0.019 |
| 3.0 μm | 1.614 | 0.012 |
| 5.0 μm | 1.725 | 0.038 |

*TABLE. 4.1 Measured phase lag for various diameters.*

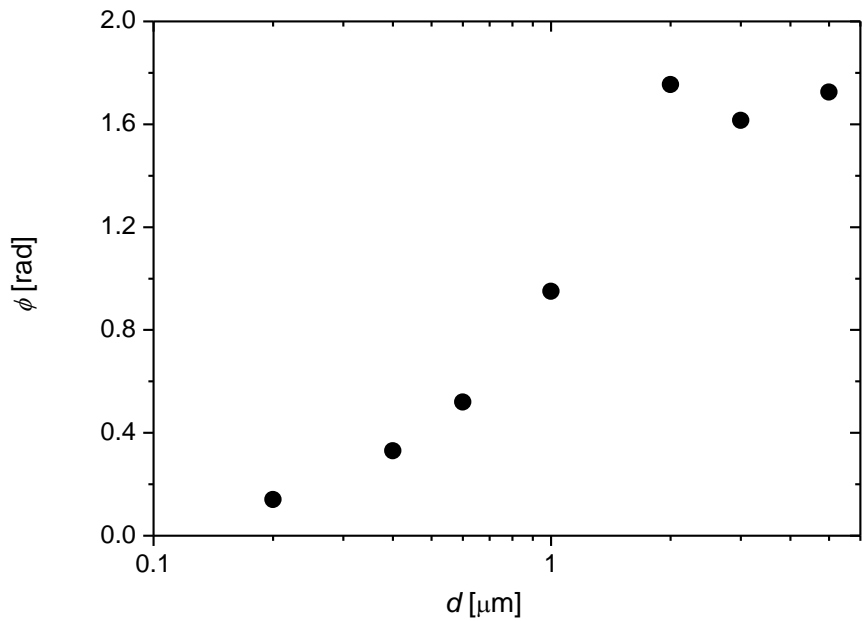

*FIG. 4.8 Measured phase delays as a function of particle diameter.*

Having determined the phase delay $\phi$, then one can give the experimental verification of the Optical Theorem by determining the modulus of amplitude function $S(0)$ (Chapter 5). The expected normalized amplitude and phase of the scattered waves using Mie theory is represented in the complex plane (figure 4.9) as a function of diameter.

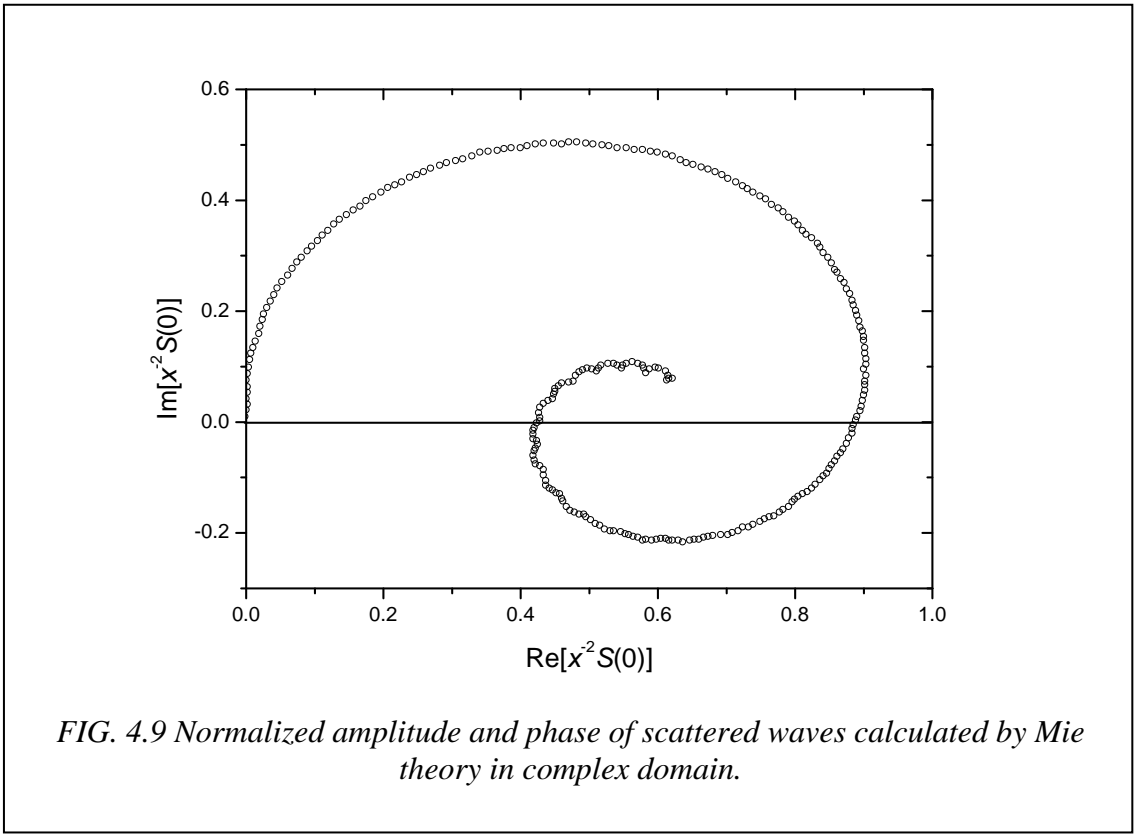

*FIG. 4.9 Normalized amplitude and phase of scattered waves calculated by Mie theory in complex domain.*

Notice the power spectrum obtained for 0.2 μm ($\Phi = 5 \times 10^{-4}$). The depth of the oscillations is smaller as compared with 0.6, 1, 3 and 5 μm samples. The reason of this result deals with one of the most striking feature of this method. As the oscillations are generated by singly scattered phase locked waves and the amount of power emerging from the sample after more than a scattering event cannot produce the Talbot effect.

In the case of the 0.2 μm ($\Phi = 5 \times 10^{-4}$) the incoming beam attenuation is 44%, determined using a calibrated solar cell. If the concentration of the sample is increased, the percentage of total single scattering events decreases accordingly and thus the depth of the oscillations is smaller when compared with other samples.

Thus, by analyzing the Talbot oscillations, one can extract the fraction of total scattered power due to single scattering alone and the method could be used to study concentrated (turbid) suspensions which are the most interest from industry (in cases dilutions are time consuming and costly), fundamental research (dilutions are not possible as it severely affect the properties of the colloid − for example emulsions, vesicles and liposomes have properties which can change in dilution).

### 4.2.3. Measurements on various concentrations of colloidal particles:

The measurements are carried out on various concentrations of 2 μm of aqueous solution of polystyrene colloidal particles. The volume fractions used and corresponding measured beam attenuations are given in the table (4.2).

| Sample | Volume Fraction ($\Phi$) | Beam Attenuations ($A$) |
|--------|--------------------------|-------------------------|
| 2 μm   | $1 \times 10^{-5}$       | 14%                     |
| 2 μm   | $5 \times 10^{-5}$       | 31%                     |
| 2 μm   | $1 \times 10^{-4}$       | 48%                     |
| 2 μm   | $3 \times 10^{-4}$       | 86%                     |
| 2 μm   | $5 \times 10^{-4}$       | 95%                     |
| 2 μm   | $7 \times 10^{-4}$       | 99%                     |

*TABLE 4.2. Various volume fractions of 2 μm samples and corresponding measured beam attenuations.*

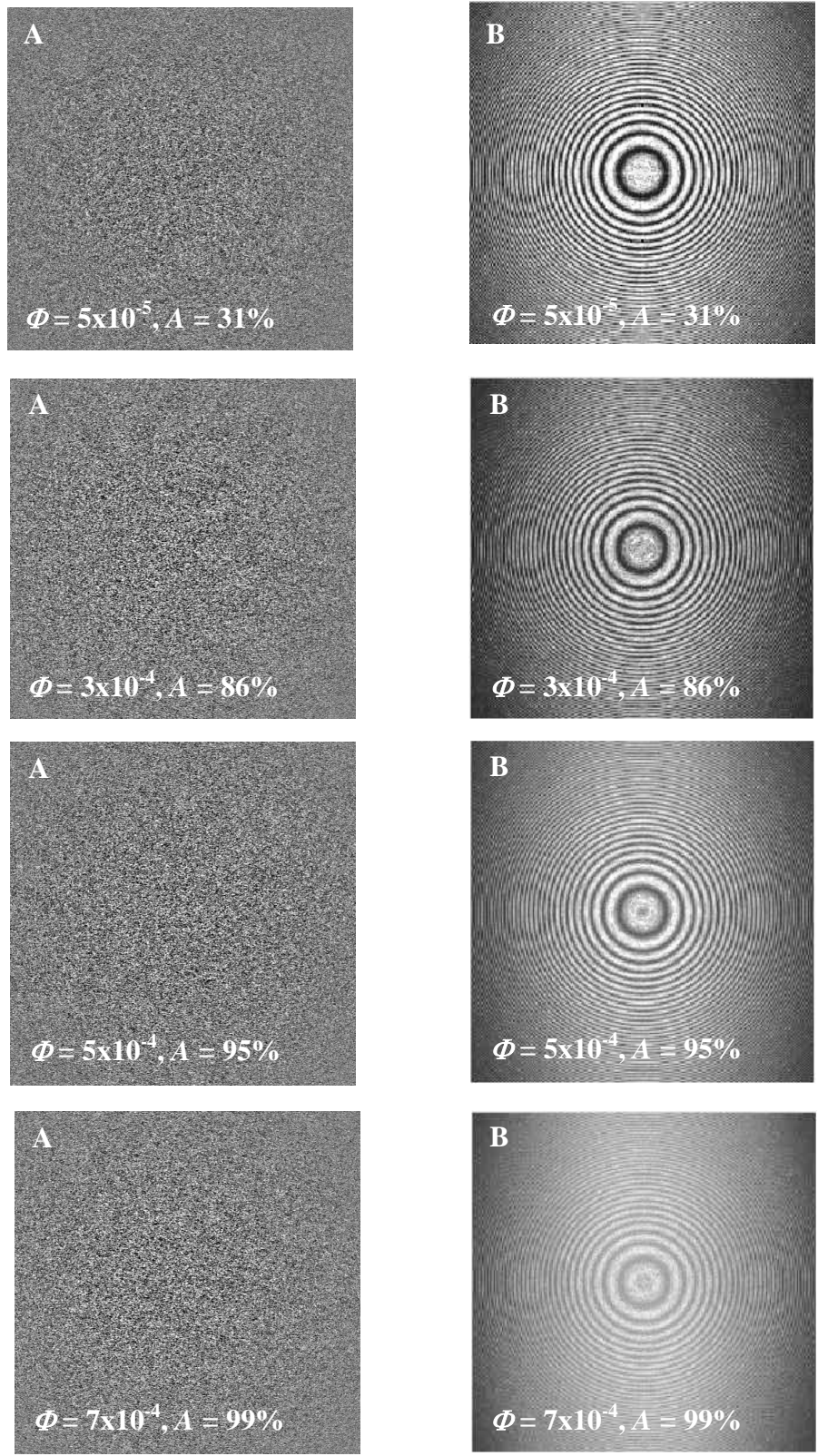

*FIG. 4.10(A) De-correlated differential image for various concentration of 2 µm samples and (B) corresponding time averaged 2-D power spectrum.*

The de-correlated differential heterodyne speckle image ($\tau$ = 100 sec) for $5 \times 10^{-5}$, $3 \times 10^{-4}$, $5 \times 10^{-4}$ and $7 \times 10^{-4}$ of 2 μm colloidal particles and corresponding time averaged 2-D power spectrums are shown in figure 4.10(A) and (B) respectively. The azimuthally averaged power spectrums are shown in figure 4.11. The graphs are plotted in relative scales.

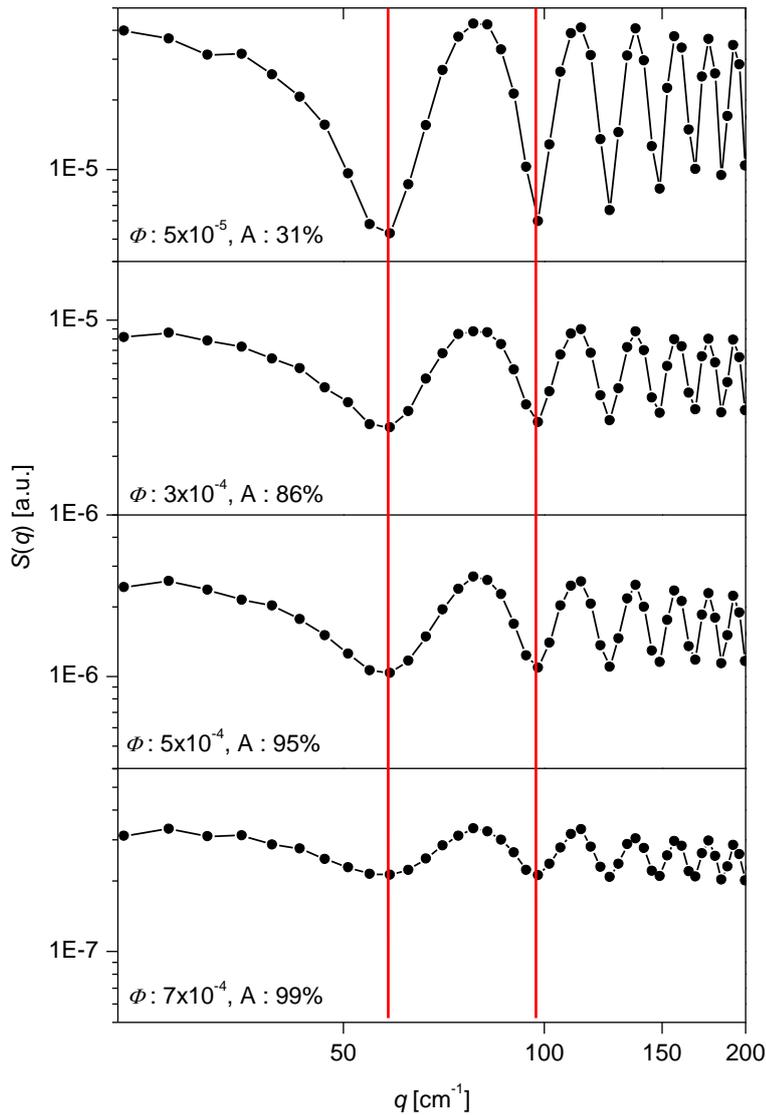

FIG. 4.11 *Azimuthally averaged power spectrum of various concentrations of 2 μm samples. Vertical red lines indicate the first and second minima of oscillations.*

Notice the azimuthally averaged power spectrum recovered for various attenuation samples. The oscillations remains fixed in $q$ for all concentrations (see the two red lines indicating the first and second minima of oscillations) as expected. The phase delays are measured by fitting eq. (4.3) and found to be same for all the concentrations within 2% error.

Also notice that the depth of the oscillations decreases as the beam attenuation increases, as expected. As mentioned earlier, the oscillations are due to single scattering alone. By analysing the depth of the oscillations, one can unambiguously identify the fraction of the total scattered power that is due to single scattering (Chapter 6).

Notice that, as it appears from the recovered power spectra of various attenuation samples that the present method works even for rampant multiple scattering (99% of incoming beam attenuations).

### 4.2.4. Dynamics:

In addition, the dynamics can be investigated by analyzing the power spectra of differences of images with short time delays as suggested in [9].

The figure 4.12(A) and (B) shows the differential heterodyne speckle images and corresponding time averaged 2-D power spectrum for time delays $\tau$ = 1, 3, 5 and 10 sec of 2 µm sample ($\Phi = 1\mathrm{x}10^{-5}$, $A$ = 12%). The azimuthally averaged power spectra for different time delays are shown in figure 4.13.

The 2-D power spectrum and corresponding azimuthally power spectrum obtained at short time delay show an evident dip at scattering wave vector ($q_x$, $q_y$) = (0, 0) and this indicates the presence of higher and higher correlations as $q \to 0$.

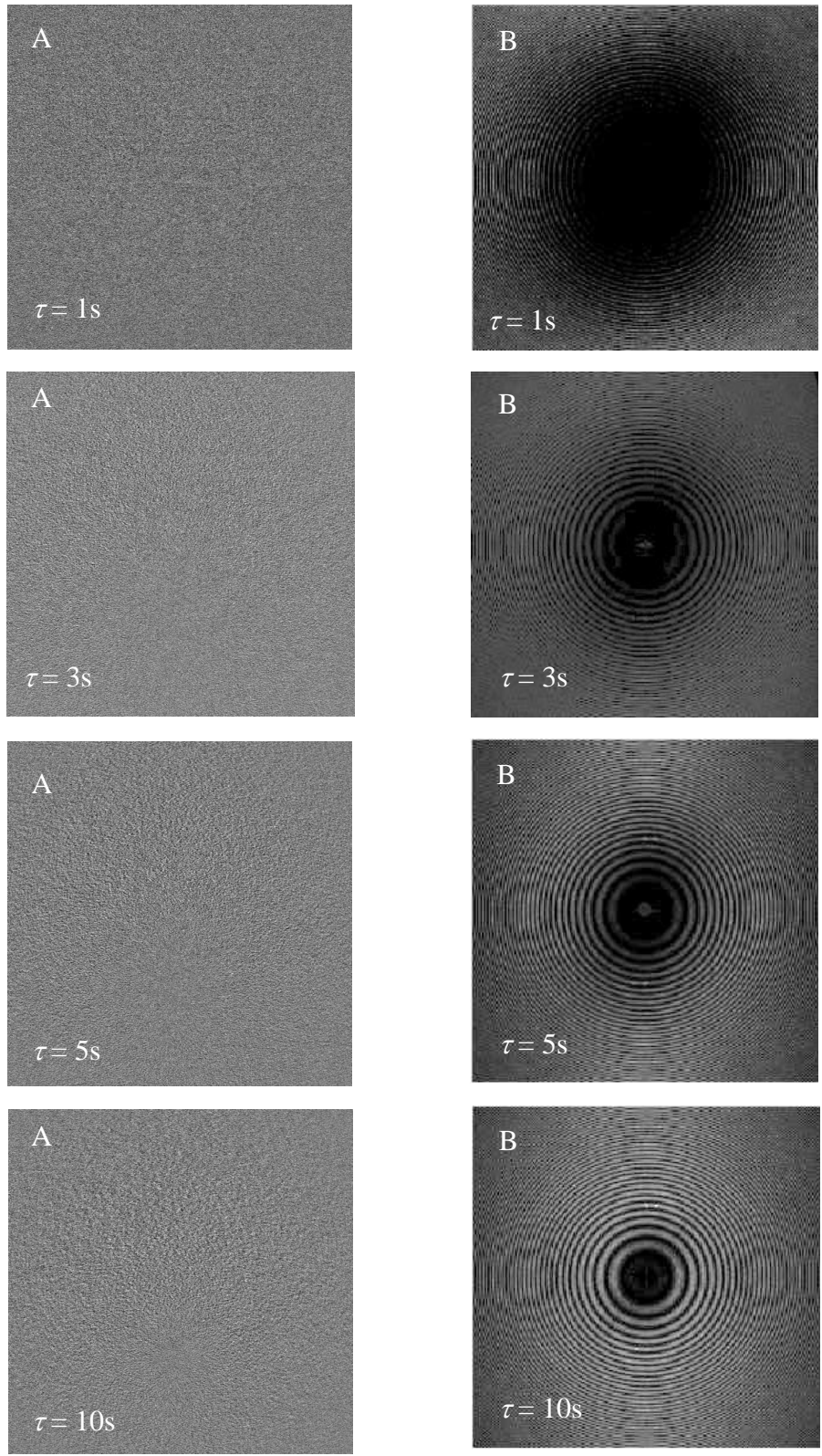

*FIG. 4.12 (A) Differential speckle images and (B) corresponding time averaged 2-D power spectrum for different time delays.*

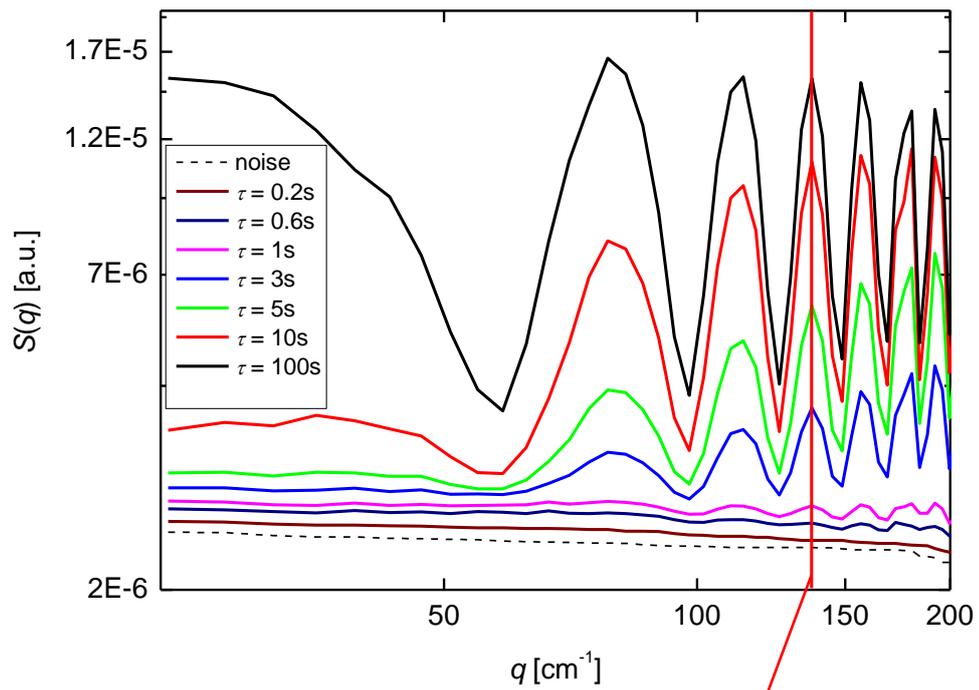

*FIG. 4.13 Azimuthally averaged power spectrum for various time delays of 2 μm (Φ = 1 x 10$^{-5}$) sample.*

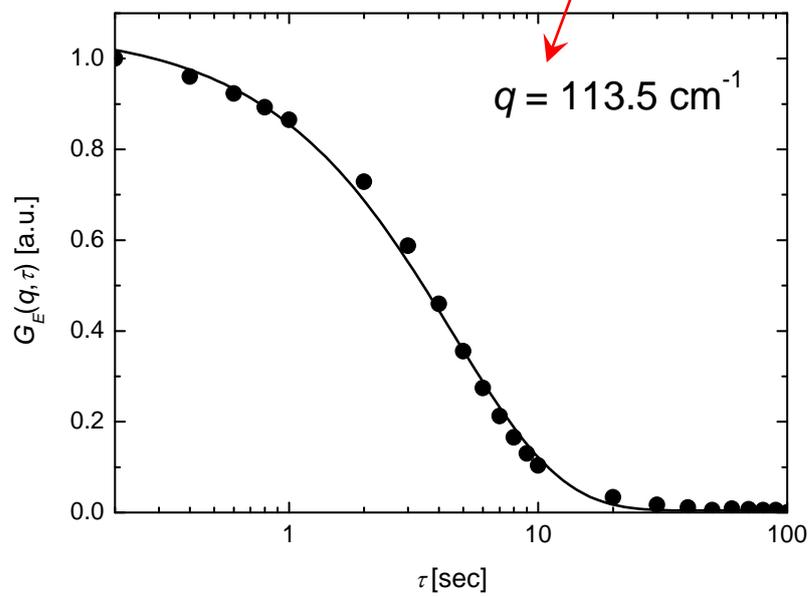

*FIG. 4.14 (●) Normalized time correlation function of 2 μm (Φ = 1 x 10$^{-5}$) sample. (——) fitting of single exponential equation.*

As the time delay increases, the spectral components of higher wave vectors are more de-correlated and if the time delay is large enough the speckles are completely de-correlated for all scattering wave vectors.

Notice that the power level grows a factor three above the noise for time delays up to 3 sec, and then the oscillations develop at any $q$ and almost complete speckle renewal at 100 s. This slow increase and saturation can be studied by calculating time correlation functions at various $q$ values. Using the method described [10], the normalized time correlation function is obtained for $q = 113.5$ cm$^{-1}$ (position of the second peak) as shown in figure 4.10. The measured correlation function is fitted with a single exponential equation with a time constant (~ 5 sec). The theoretical diffusive time constant at $q = 113.5$ cm$^{-1}$ is 3 x 10$^5$ sec.

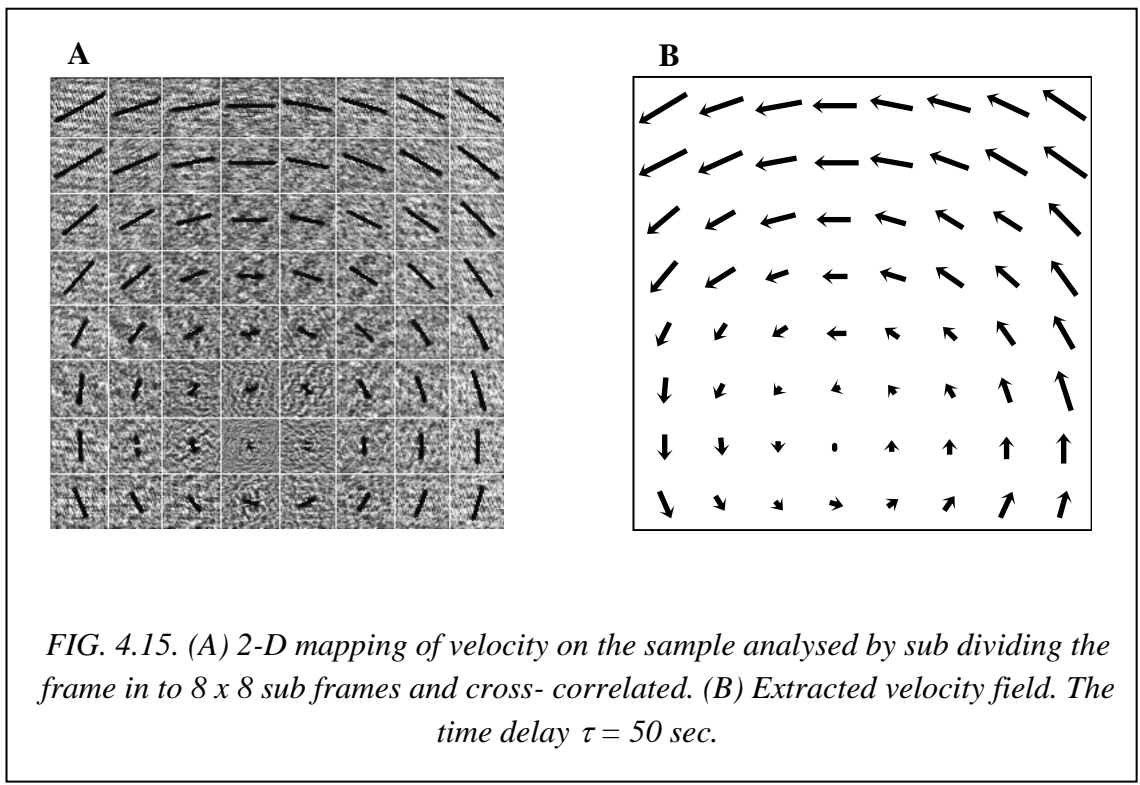

*FIG. 4.15. (A) 2-D mapping of velocity on the sample analysed by sub dividing the frame in to 8 x 8 sub frames and cross- correlated. (B) Extracted velocity field. The time delay $\tau = 50$ sec.*

The measured dynamics is found to be comparatively fast compared to what expected for Brownian motion. This is due to a slow macroscopic convective motion as it can be verified by dividing the frame into many sub frames and cross correlating each sub frame. This provides the velocity vector in each portion of the sample [11, 12]. The 2-D mapping of velocity is shown in figure 4.15 (A) and (B).

Much more interesting from a dynamical point of view is the case of strong concentration and strong beam attenuation, a regime where multiple scattering is rampant. The figure 4.16 shows the azimuthally averaged power spectrum for various time delays of 2 μm sample ($\Phi = 7 \times 10^{-4}$, $A = 99\%$).

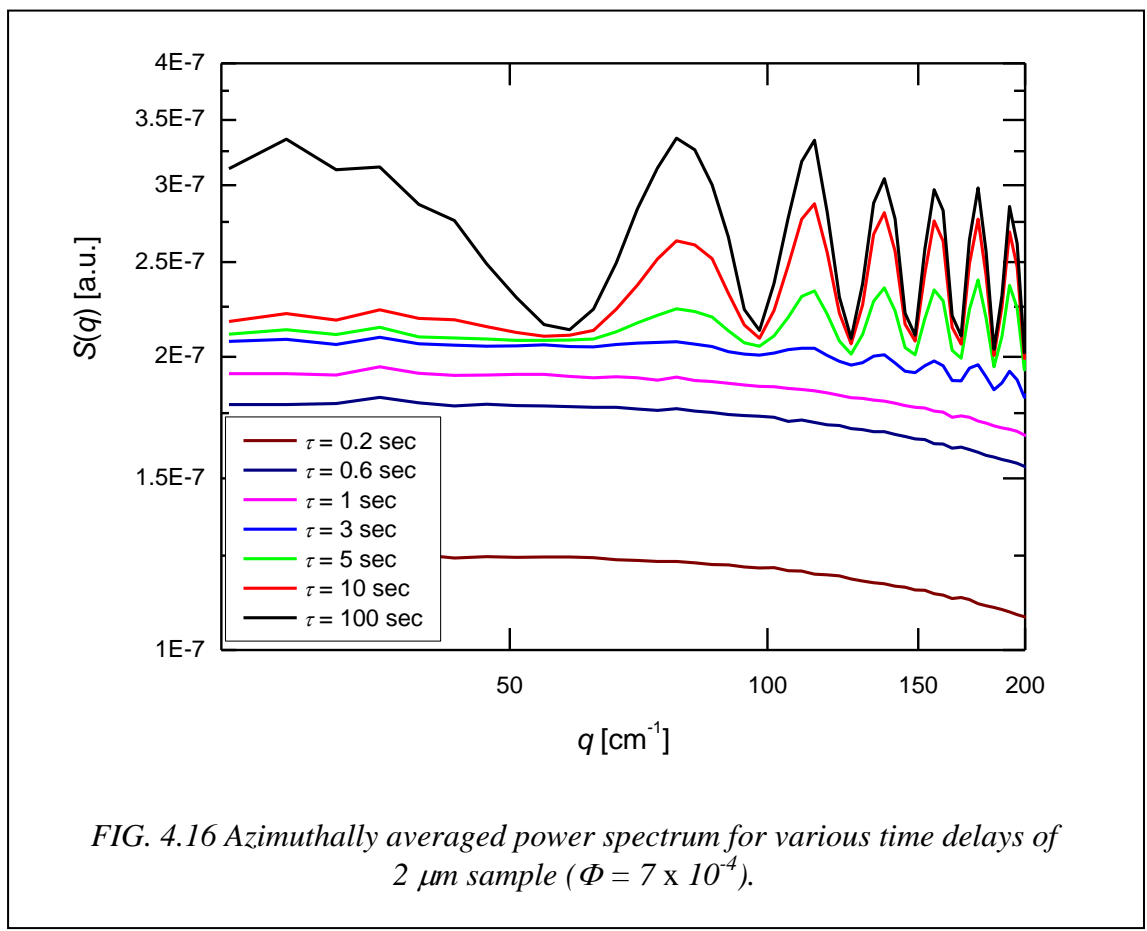

*FIG. 4.16 Azimuthally averaged power spectrum for various time delays of 2 μm sample ($\Phi = 7 \times 10^{-4}$).*

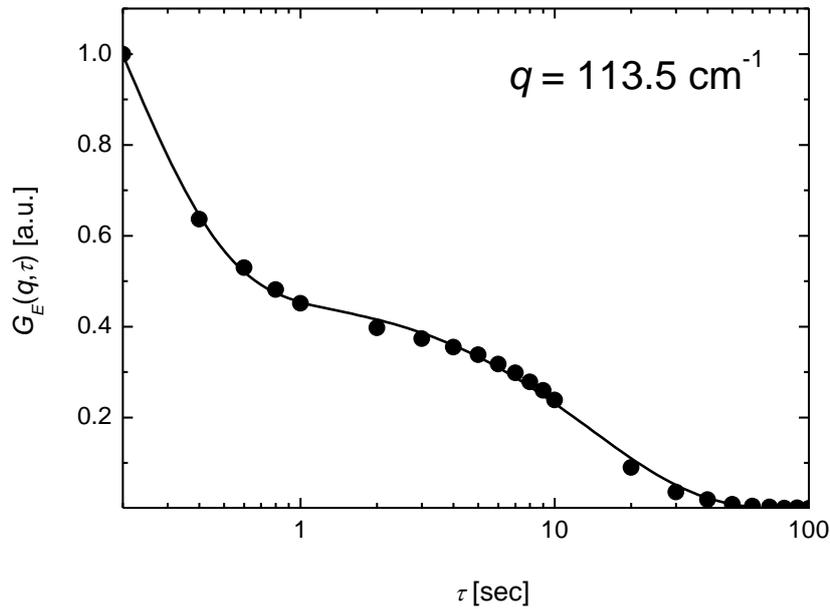

FIG. 4.17 (●) *Normalized time correlation function of 2 μm (Φ = 7 x 10⁻⁴) sample. (——) fitting of double exponential equation.*

One can notice the rapid increase in the power level at short times, say less than 1 sec. All the curves do not show any oscillation, and the shape of the curves does not change (the relative power levels are the same in figure 4.13 and 4.16). At longer time delays ($\tau = 3$ sec), the oscillations begin to develop and they saturate to a constant profile as in dilute concentration.

As previously pointed out, the oscillations are due to single scattering alone. As a consequence, the smooth, oscillation free curves at short time delays are due to multiple scattering. The analysis of the time correlation function does indeed show this, as the function shows that it can be decomposed into a fast decay (due to multiple scattering) and a slow decay (due to convection motion) as shown in figure 4.17. The former decay corresponds to power spectra without oscillations, and then it is certainly due to multiply scattered waves. The later is related to oscillating power spectra, and it is then attributed to singly scattered fields.

## 4.3. Perspectives

It is worth notice that the Near Field based low angle scattering method is capable to do that previously considered as impossible, namely measure the properties of the forward scattered wave. This is really impossible by any classical light scattering methods as the scattering process is conceived as Bragg reflection from a three dimensional random grating. By contrast, the present method exploits the so called Raman-Nath scattering regime; the phase relationship between the symmetrically scattered waves preserves the information about the phase delay.

Having determined the phase delay of various diameters of mono dispersed colloidal particles, the experimental verification of the Optical Theorem could be attempted (Chapter 5). The measurements on various concentrations of mono dispersed colloidal particles reveals that the method could provide the contribution due to single scattering in turbid media by analyzing the depth of the oscillations (Chapter 6). Measuring phase delay of scattered waves in principle could measure the optical thickness of the scatteres, a quantity of interest in phase transition of colloidal studies. In chapter 7 the method is used for studying aggregating colloids.

# Chapter 5

# Verification of Optical Theorem

# Chapter 5

# Verification of Optical Theorem

The Optical Theorem (OT) has a long history. The occurrence of the OT in electromagnetic theory begins more than hundred years ago. The applications of the theorem are found in quantum mechanical and acoustic scattering [4].

## 5.1. Optical Theorem

Let us consider a linearly polarized plane wave of light (incident wave) impinging on a single particle *P*, of arbitrary shape and composition. Beyond the particle, due to the interactions between the particle and the incident wave, a fraction of the power of incident wave is removed and redirected. In addition, a fraction of the power can also be absorbed by the particle composition. The absorption accounts the power directed in to the particle interior.

The figure 5.1 shows the reduction of the power collected by the detector placed in forward direction due to the particle scattering and absorption. The reduction of power in the forward direction is often referred to as extinction effect [13].

Let the total power scattered in all directions be equal to the power of the incident wave falling on the area $C_{sca}$. Likewise, the energy absorbed inside the particle be equal to the power incident on the area $C_{abs}$, and the power removed from the original incident wave be equal to the power incident on the area $C_{ext}$.

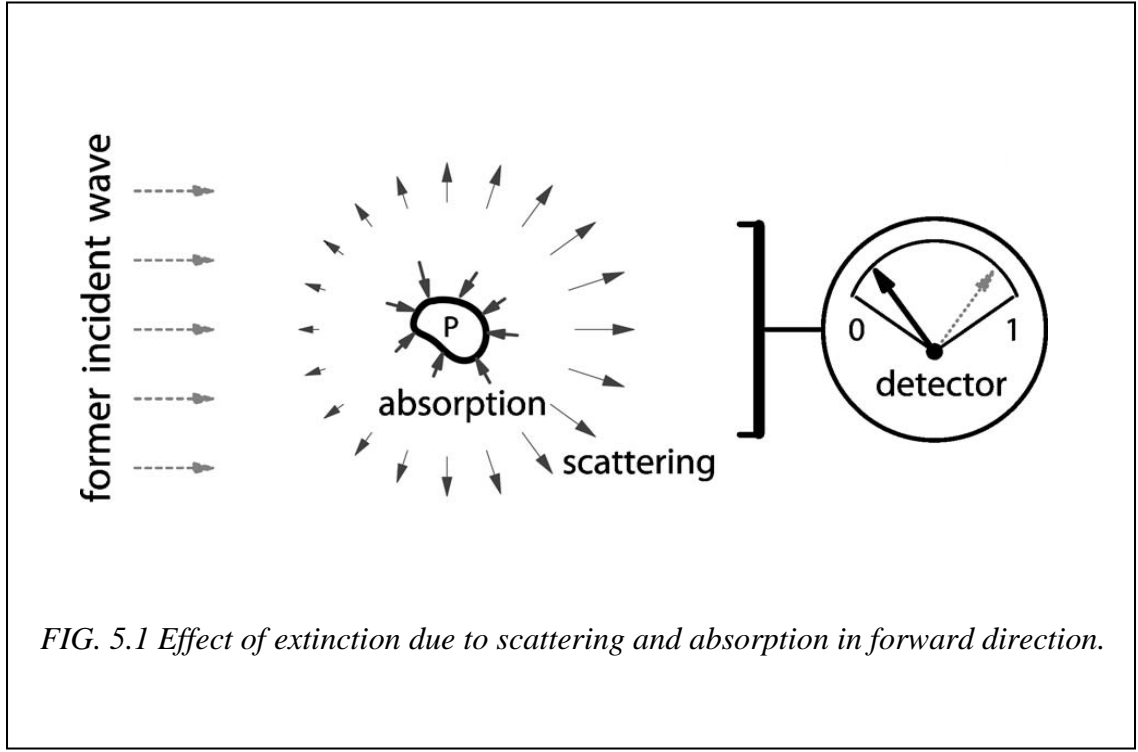

*FIG. 5.1 Effect of extinction due to scattering and absorption in forward direction.*

Using the law of conservation, then

$$C_{ext} = C_{sca} + C_{abs} \qquad \text{eq. (5.1)}$$

The quantities $C_{sca}$, $C_{abs}$, $C_{ext}$ are the cross-sections of the particle for extinction, scattering and absorption respectively. For non-absorbing particles,

$$C_{ext} = C_{sca} \qquad \text{eq. (5.2)}$$

The Optical Theorem states that to obtain the total extinction cross-section (including both scattering and absorption) the only thing needed is the knowledge of the complex amplitude function $S(0) = |S(0)| \exp(i\,\phi)$, where $\phi$ is the phase delay and $|S(0)|$ is the modulus of amplitude function. Explicitly, the Optical Theorem can be expressed by the formula

$$C_{ext} = \frac{4\pi}{k^2} \operatorname{Re}\{S(0)\} \qquad \text{eq. (5.3)}$$

This is the fundamental extinction formula, where $k = 2\pi/\lambda$ ($\lambda$ being the wavelength of incoming wave).

For a cloud of many independent scattering particles randomly distributed, then each scattered wave is characterized by the amplitude functions $S_i(0)$. The index $i$ denote any individual particle. Repeating the argument for a cloud of particles,

$$S(0) = \sum_i S_i(0) \qquad \text{eq. (5.4)}$$

and as a consequence,

$$C_{ext} = \sum_i C_{i,ext} \qquad \text{eq. (5.5)}$$

The physical understanding of the OT relies on the interpretation of extinction as caused by the interference of the incident wave and the scattered waves in only zero-angle forward direction. This interference between the incident and scattered wave accounts for the $C_{ext}$ via a phase relation between these two waves. Thus, in order to verify the OT, the only thing needed is the knowledge of phase relation between the incident and scattered wave (phase delay). Although it is not widely recognized, the Mie theory predicts this phase change between the incoming plane wave and the scattering spherical wave at zero angles. For refractive particles the change occurs when the particle optical thickness (the refractive index multiplied by the path length of light within the particle) varies between a small fraction of $\lambda$ up to more or less $\lambda$, the total change becoming asymptotically equal to $\pi/2$ [3].

In an authoritative paper where the possibility of an experimental verification of OT is discussed [3, 4], it was pointed out that a proof could hardly come from an optics experiment. If one conceives the scattering from an assembly of particles as Bragg reflection from a 3-D random grating, then this statement is correct. Accordingly, the scattered waves at any angle then behave as a circular Gaussian process, and phases are equally distributed in the interval [5], therefore giving no cue on the behaviour of the wave scattered at exactly zero angle.

In the succeeding sections, the complex amplitude function is determined by measuring the phase of the scattered waves for suspensions of various diameters of mono disperse colloidal particles is presented using Near Field scattering method and the modulus of the scattering amplitudes are extracted by taking the ratio of the variance and the main beam intensity. Knowing Phase delay and modulus of scattered wave, the Optical Theorem can be verified by calculating the extinction cross-section eq. (5.3).

## 5.2. Phase Delay Measurements Using Near Field Scattering Method

The near field scattering technique works for so small angles such that the Raman-Nath scattering conditions are full filled (Chapter 2). In this regime, the scattering is due to 2-D phase gratings and any scattering event always generates two phase locked scattered waves at symmetric angles and it is the three wave interference that generates this peculiar oscillatory modulation, related to the so called Talbot effect [8]. If $A_o$ is the amplitude of the incident wave, $S(\theta)$ is the scattering amplitude function of the scattering particle, $\lambda$ the radiation wavelength, $k = 2\pi/\lambda$. Thus the corresponding interference pattern with the transmitted beam on the sensor at a given time $t$ is given by

$$I(x,y) = |A_o|^2 + \frac{2A_o|S(0)|}{kz} \cos\left[k\frac{x^2+y^2}{2z} - \phi + \frac{\pi}{2}\right] \qquad \text{eq. (5.6)}$$

Here the interference pattern is evaluated at a distance $z$ along in the forward direction, such that the paraxial approximation holds and $S(\theta) \sim S(0)$, and the phase delay $\phi = \text{Arg}[S(0)]$ is introduced, determining the fractional order at the centre of the fringe pattern. Although the intensity distribution generated by a collection of many scatterers appears as a completely stochastic signal, the fact that they are made up by circular interference fringes becomes evident when the spatial power spectrum $S(q_x, q_y)$ of the intensity distribution $I(x, y)$ is taken.

$$S(q_x, q_y) = \frac{4\pi^2}{k} |S(0)|^2 \sin^2\left[\frac{q^2 z}{2k} - \phi\right] \qquad \text{eq. (5.7)}$$

Both the functions eq. (5.6) and eq. (5.7) have an infinite sequence of oscillations, and they possess a unique property, namely that all the power under the 2$i$-th oscillation of $S(q_x, q_y)$ comes from the $i$-th fringe of $I(x, y)$ alone.

| Sample | Volume Fraction ($\Phi$) |
|---|---|
| 0.2 μm | 5 x $10^{-4}$ |
| 0.6 μm | 5 x $10^{-5}$ |
| 1 μm | 5 x $10^{-5}$ |
| 2 μm | 5 x $10^{-5}$ |
| 3 μm | 5 x $10^{-5}$ |
| 5 μm | 5 x $10^{-5}$ |

*TABLE 5.1. Various diameters and their volume fractions.*

The figure 5.2 shows the spatial power spectrum obtained for mono disperse polystyrene spheres of various diameters as listed in table (5.1). In figure 5.3, the spatial power spectrum of 1 μm ($\Phi = 5 \times 10^{-5}$) is fitted with eq. (5.7) to measure phase delay. Similarly the phase delays are measured by fitting eq. (5.7) for diameters (0.2, 0.6, 2, 3 and 5 μm).

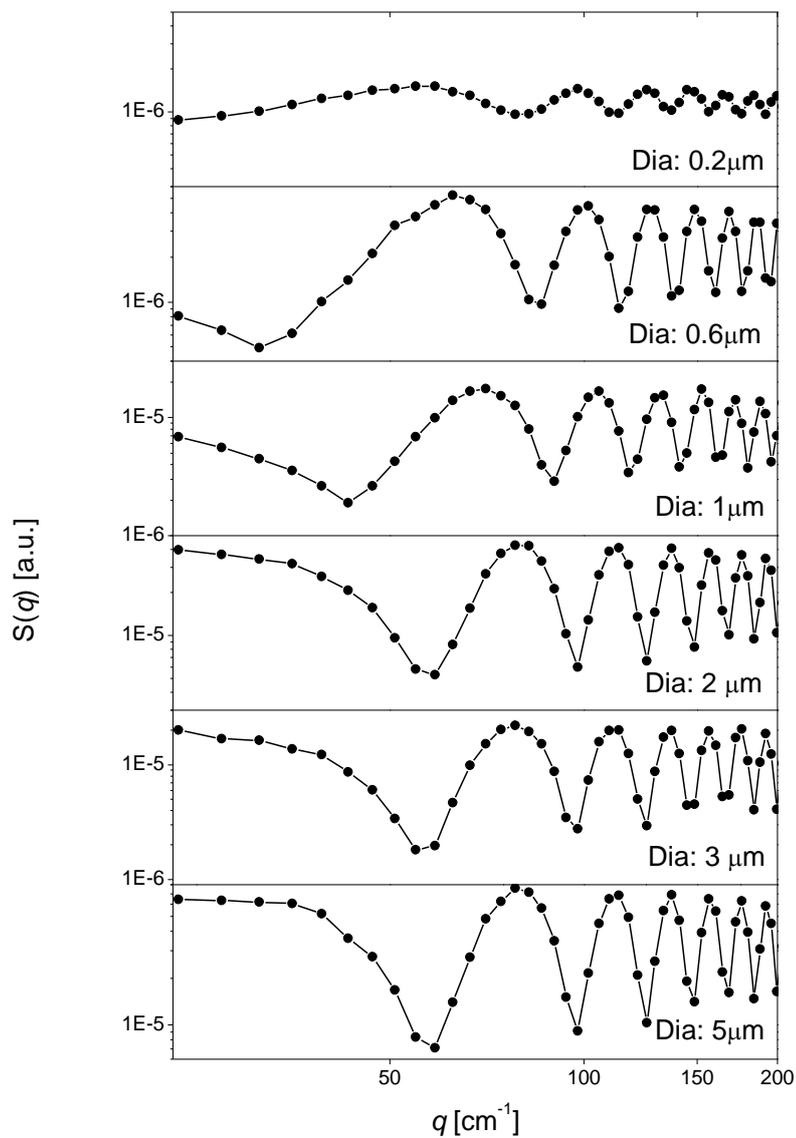

*FIG. 5.2 Spatial power spectrum obtained for various diameters.*

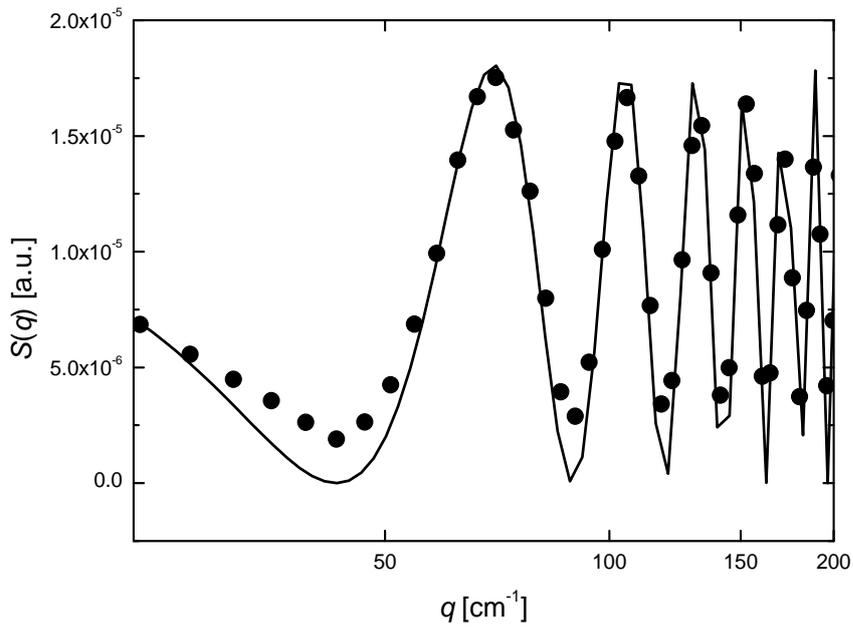

*FIG. 5.3 (●) Power spectrum for 1 μm ($\Phi = 5 \times 10^{-5}$), (——) best fitting of eq. (5.7).*

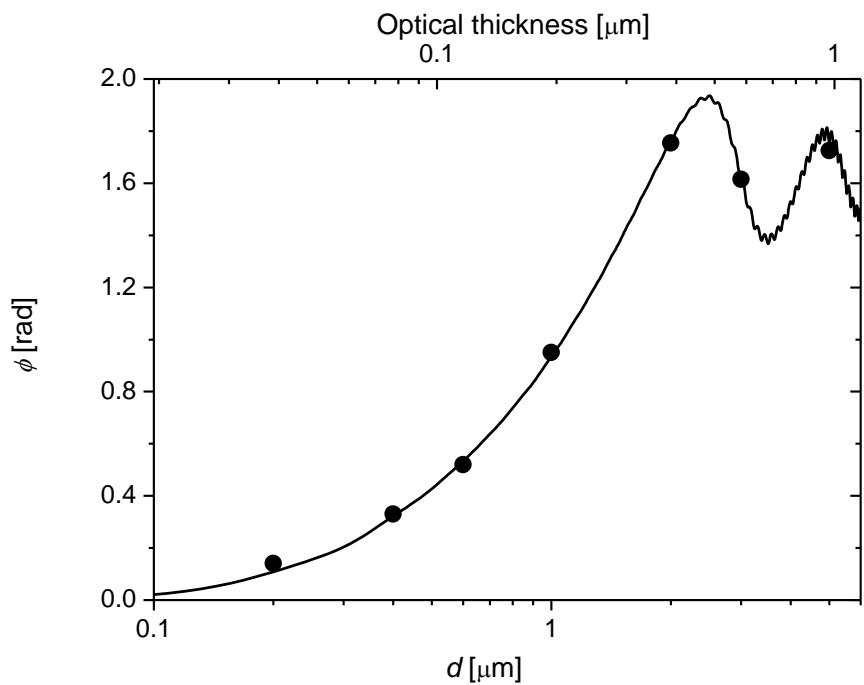

*FIG. 5.4 (●) Phase delay as a function of particle diameter and optical thickness, (——) theoretical value obtained from Mie theory.*

The phase delay is compared to the expected value as obtained by using the Mie theory (a series expansion). The experimental results for phase delays are shown in figure 5.4 as a function of particle diameter and optical thickness, together with the theoretical values calculated using Mie theory.

When comparing the present method with the any classical static scattering methods, it must be emphasized that this is the only method that measures the phase delay between the incoming wave and the scattered wave.

## 5.3. Modulus of *S*(0) Measurements Using Near Field Scattering Method

Since the scattering method exploits a self referencing (heterodyne) scheme, the modulus of the scattering amplitude *S*(0) can be extracted from the ratio of the variance and the main beam intensity. In the following a method for evaluating *S*(0) from experimental data is presented.

### 5.3.1 First order statistics of intensity:

The variance of the heterodyne signal can be related to the amplitude *S*(0) under known experimental conditions.

The variance of the heterodyne signal is characterized by first order intensity distribution as Gaussian function with average $A_T^2$ [14] is given as

$$V_I = 4\sigma^2(A_T^2 + \sigma^2) \approx 4\sigma^2 A_T^2 \qquad \text{eq. (5.8)}$$

where

$$2\sigma^2 = \sum_{j=1}^{N} a_j^2 = N\langle a_j^2 \rangle \qquad \text{eq. (5.9)}$$

$$a_j(x,y) = |S(0)| \frac{A_j}{k_o |r - r_j|} \qquad \text{eq. (5.10)}$$

$A_T$ is the amplitude of the transmitted field, $A_j = A(r_j)$ is the amplitude of the field emerging from the *j-th* scatterer placed at position $r_j = (x_j, y_j, z_j)$, $N$ is the number of particles. As a consequence, the variance of the heterodyne speckle intensity is given as

$$V_I = 2N \sum_{j=1}^{N} |S(0)|^2 \frac{A_j^2 A_T^2}{k_o^2 z^2} \qquad \text{eq. (5.11)}$$

where the approximation $|r - r_j| = z$ used for any $j$.

The dependence on distance z, together with the fact that the sample is illuminated by a Gaussian beam profile must be accounted for. By limiting the analysis at the region of the sensor close to the optical axis, the amplitude of the scattered waves depends only on the distance from the optical axis, $r = z\,\theta$. Any distance $r$ is associated with a value $I(r)$ for the intensity of the incoming beam:

$$|A(r)|^2 = I(r) = I_m \exp(-r^2 / r_0^2) \qquad \text{eq. (5.12)}$$

where $r_0$ is the distance at which the beam intensity drops to a value *1/e* of the maximum, $I_m$. For a given annular region of the sample between a distance $r$ and $r + dr$ from the optical axis, the variance of the speckle field is given by:

$$dV_I = 2n2\pi r dr L |S(0)|^2 \frac{A_T^2 A_m^2}{k_0^2 z^2} \exp\left(-r^2/r_0^2\right) \qquad \text{eq. (5.13)}$$

where *n* is the number density of the monomers and *L* the sample thickness. In principle this should be integrated from $r = 0$ up infinity, that means either up to $r = r_{max}$ large enough to guarantee the correct result, or to the physical boundary of the sample. The expression for the variance then becomes:

$$V_I = 4n\pi L |S(0)|^2 \frac{A_T^2 A_m^2}{k_0^2 z^2} \int_0^{r_{max}} \exp\left(r^2/r_0^2\right) r dr \qquad \text{eq. (5.12)}$$

Once the variance is measured from the heterodyne speckle field, this result allows to measure to get the scattering amplitude *|S(0)|*. The independent measure of the phase lag *ϕ* fully determines the complex amplitude, *S(0)* and the OT can be used to get the extinction cross sections.

### 5.3.2 Data reduction scheme to measure variance

Using the data reduction scheme, the normalized intensity distribution is given by

$$I(x, y) = \frac{I(x, y) - A_T^2}{A_T^2} \qquad \text{eq. (5.15)}$$

which gives a normalized variance $V = V_I / (A_T^2 A_0^2)$. The result is then independent of the incoming beam amplitude, *A(r)* and the transmitted beam amplitude, $A_T$. Notice that the difference of frames can also be used, and the speckle variance is obtained from the normalized intensity distribution given by

$$I(x,y) = \frac{I_1(x,y) - I_2(x,y)}{I_1(x,y) + I_2(x,y)} \qquad \text{eq. (5.16)}$$

which gives a value for the variance a factor 2 smaller than before. If the extinction is very small, that is $A_T = A_m$, the normalized variance can be reduced to

$$V = \frac{V_I}{A_T^4} \qquad \text{eq. (5.17)}$$

### 5.3.3 Samples with high turbidity:

If huge extinction is present, then the transmitted beam intensity is depressed according to the Lambert-Beer law $A_T^2 = A_i^2 \, exp(-\gamma \, L)$, where $\gamma$ is the extinction coefficient and $A_i$ is the incoming beam amplitude. Similarly, the beam falling onto the scatterer placed at a depth z within the sample will be reduced according to $A^2 = A_i^2 \, exp(-\gamma \, z)$. Also the scattered wave suffers the extinction, and in the forward direction it causes a decrease in the intensity given by a further factor $exp[-\gamma \, (L-z)]$. As a result, the emerging scattered wave is depressed by the same extinction factor as the transmitted beam is, if compared to the scattered wave in the case where no extinction is present (eq.5.17) will then provide again the correct information for evaluating *|S(0)|*, provided that only the singly scattered waves are accounted for.

The analysis of the power spectrum of the speckle field is then precious, since the Talbot oscillations are genuinely due to the singly scattered radiation interfering with the transmitted beam. Furthermore, when multiple scattering is present, the Talbot oscillations are limited in amplitude, the minima lying on a plateau given by power spectral density due to the multiply scattered waves. As a consequence the single scattering contribution to the variance can be easily obtained by taking the variance of the overall speckle field multiplied by the ratio between half the oscillation amplitude and the overall power spectral density. This procedure allows to get rid of the contributions given by the multiple scattering.

## 5.4 Extinction Cross section

Once the variance is measured from the normalized intensity distribution *of* the speckle field, this result allows to get the scattering amplitude $|S(0)|$.

The independent measure of the phase lag $\phi$ fully determines the complex amplitude,

$$S(0) = |S(0)| \exp(i\,\phi) \qquad \text{eq. (5.18)}$$

Then Optical Theorem can be used to get the extinction cross section:

$$C_{ext} = \frac{4\pi}{k^2} \text{Re}\{S(0)\} \qquad \text{eq. (5.19)}$$

and compared to the Mie exact solution for the light scattering from spheres as shown in figure 5.5. A better comparison can be made by calculating the expected amplitude and phase of the scattered waves and by representing those in the complex plane (see figure (5.6)). Good agreement is reported.

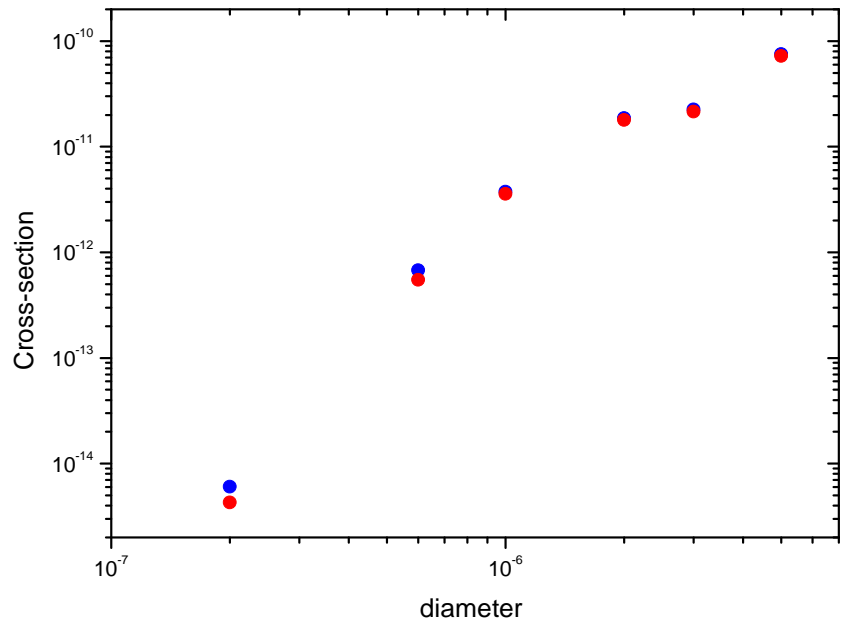

*FIG 5.5 Cross-section as a function of diameter. (•) Experimental and (•) Calculated.*

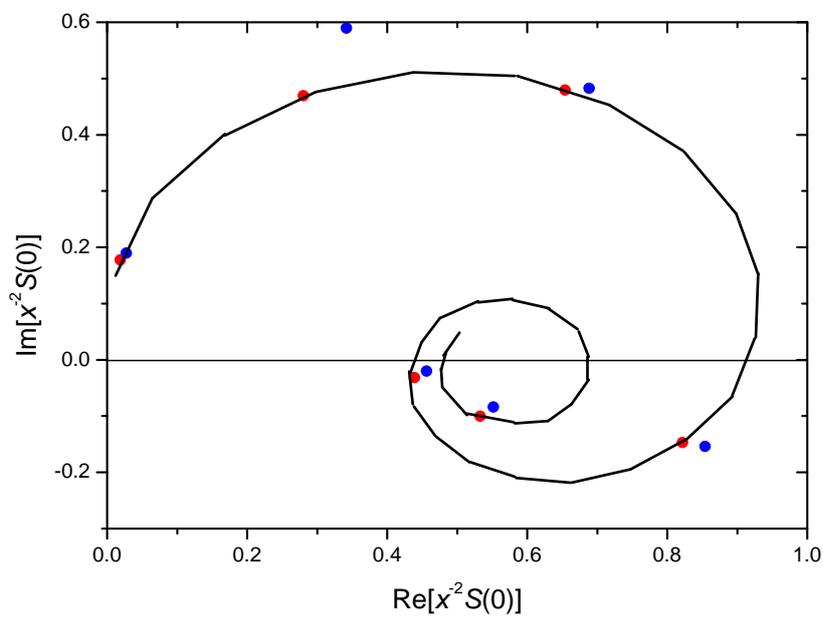

*FIG. 5.6 Normalized amplitude and phase of scattered waves in complex domain. (•) Experimental and (•) Mie theory.*

## 5.5 Conclusion

The Near Field scattering method thus measures the phase of the scattered waves and the modulus of the scattering amplitude and the celebrated Optical Theorem is experimentally verified for first time in optics.

*Chapter 6*

*Measure of Single Scattering in Turbid Colloidal Suspensions*

# Chapter 6

# Measure of Single Scattering in Turbid Colloidal Suspensions

Light scattering methods are well established techniques to characterize the sub-micron systems such as colloids, polymers, gels and biological macromolecules etc [1,2]. However, both Static light scattering and Dynamic light scattering techniques assume that the detected photons have undergone only one scattering event i.e., the multiple scattering is negligible.

Under most conditions, this assumption is valid only for very dilute suspensions (beam attenuations $A < 10\%$). However, much of the interest from industry, and increasingly of fundamental research, is in the study of turbid suspensions exceeding more than $A > 50\%$. Thus in order to characterize a suspension using light scattering, it must be diluted. In many cases the dilutions are time consuming and costly, while in others dilution is not possible as it severely affects the properties of the suspensions (emulsions, vesicles, etc). In recent years, the study of turbid suspensions has been facilitated by the development of new techniques [15].

In this chapter, the study on various concentration of mono disperse colloidal particles (up to 99 % beam attenuation) using Near Field based low angle scattering method are presented. The study reveals that the method can identify the fraction of total scattering power due to single scattering alone.

## 6.1. Measurements on Various Concentrations of Colloidal Particles

In practice, when the concentration of the suspension is increased, then the amount of power emerging from the sample have more than one scattering events (multiple scattering). There are few methods to detect single scattering in multiple scattering suspensions but they are limited to specific applications [15].

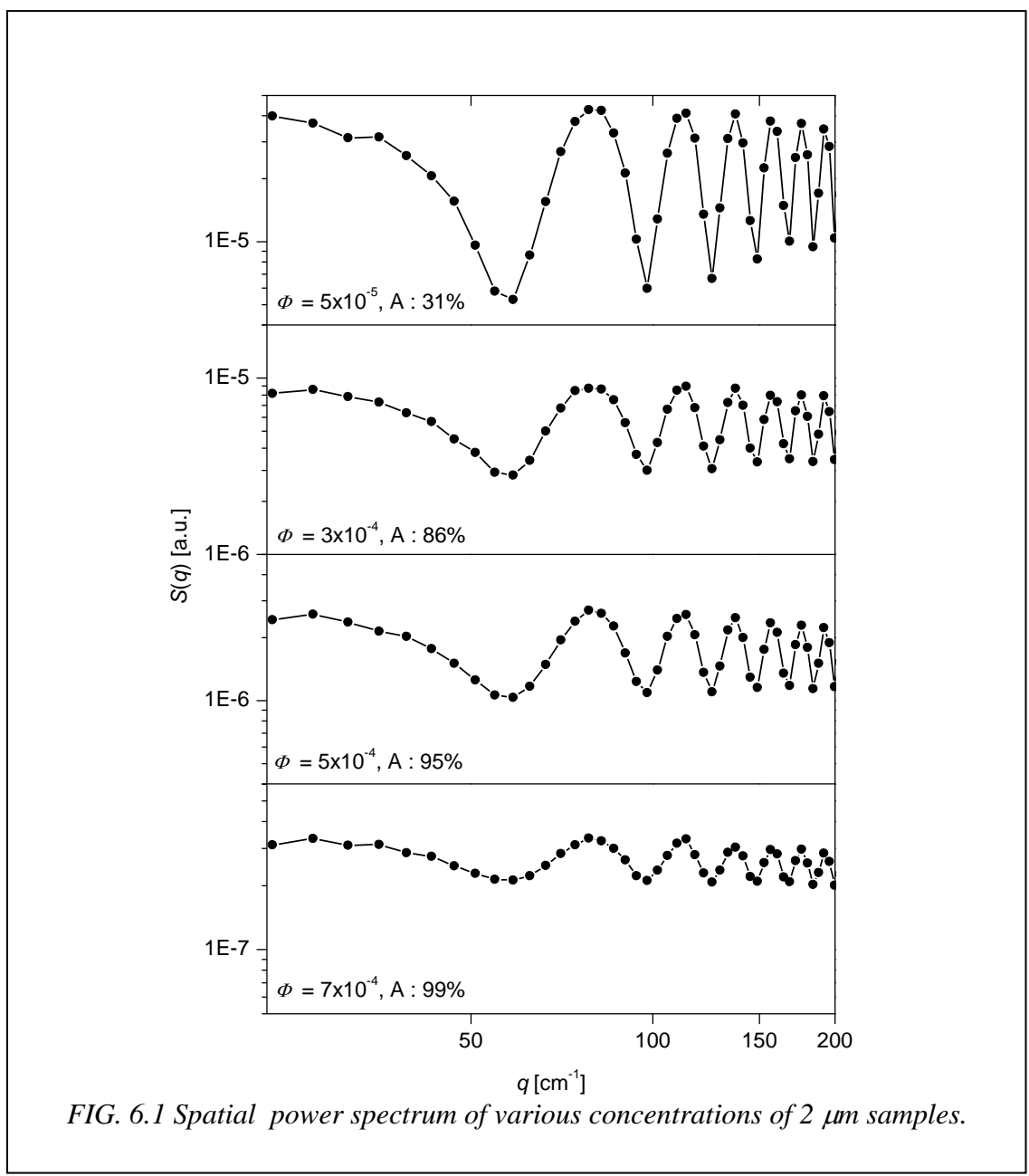

*FIG. 6.1 Spatial power spectrum of various concentrations of 2 μm samples.*

The measurements using Near Field based low angle scattering method exhibits Talbot like oscillations in the spatial power spectrum. The oscillations are due to phase locked three waves interference as discussed in Chapter 4, namely due to single scattering only.

The figure 6.1 shows the spatial power spectra obtained for various beam attenuation of 2 µm particles. Notice the presence of Talbot like oscillations even for 99 % beam attenuated sample. The depth of the oscillations is much smaller compared to other samples as expected. This is because the amount of power emerging from the sample after more than one scattering event is not involved in the Talbot effect. As the concentration (beam attenuation) increased, the percentage of total single scattering events decreases accordingly.

By analyzing the Talbot oscillations the fraction of power due to single scattering can be recovered.

## 6.2. Extraction of Single Scattering Fraction

The fraction of single scattering can be obtained as

Single scattering fraction = Single scattering / Total scattering            eq. (6.1)

As discussed already, the contribution of single scattering can be recovered from the depth of the oscillations and the total scattering can be recovered from the 2-D integration of spatial power spectrum.

In the data reduction scheme, the static spatial power spectrum (Chapter 3) is given by

$$S(q) = 4\left|E_O(q) * E_S^*(q,t)\right|^2 +$$
$$4\,\mathrm{Re}\left\{E_O(q) * E_S^*(-q,t) \cdot E_O(q) * E_O^*(-q,t) + E_O^*(-q) * E_S(q,t) \cdot E_0^*(-q) * E_S(q,t)\right\}$$

eq. (6.2)

where, $E_s$ is the total scattering field which includes the single and multiple scattering, $E_o$ is the transmitted filed. In eq. (6.2), the first term relates the 2-D integration of spatial power and the second term relates the oscillation terms (maxima and minima).

If, the maxima of oscillation are denoted by $S_{max}$ and the minima of oscillation are denoted by $S_{min}$, then the single scattering and total scattering power can be recovered (see figure 6.2).

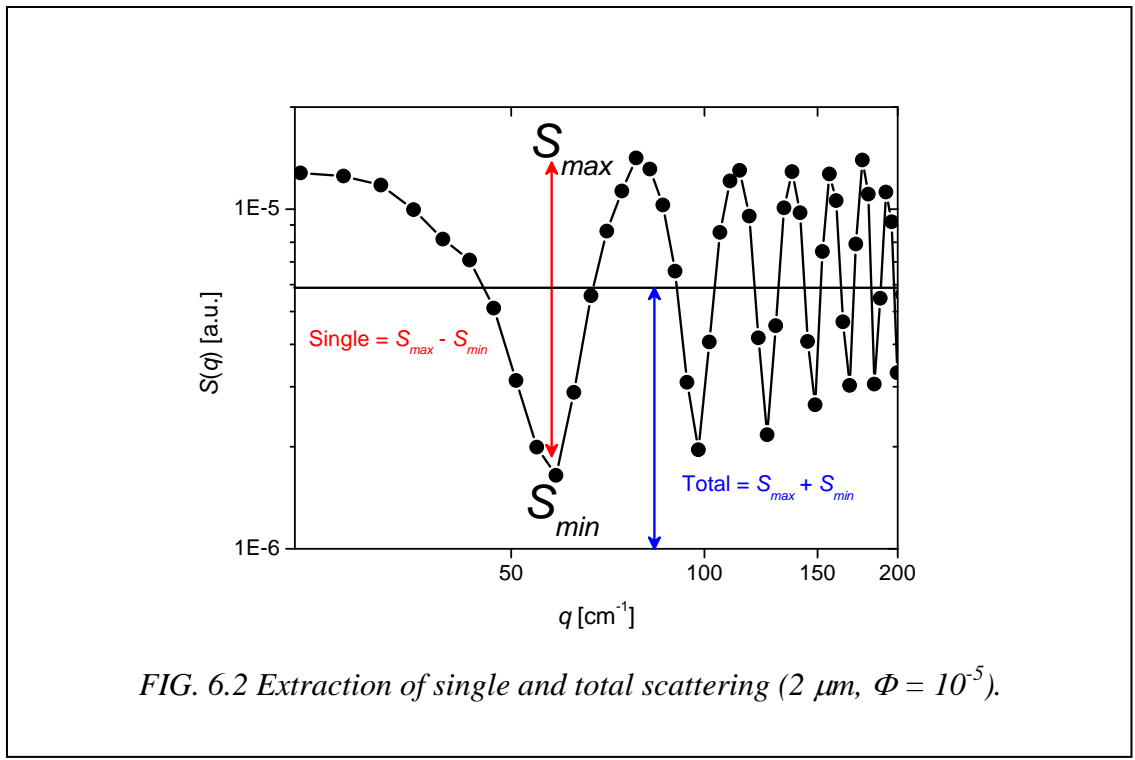

FIG. 6.2 Extraction of single and total scattering (2 $\mu$m, $\Phi = 10^{-5}$).

Thus, the depth of the oscillation can be recovered by

$$S_{max} - S_{min} \qquad \text{eq. (6.3)}$$

and the total power can be recovered by

$$S_{max} + S_{min} \qquad \text{eq. (6.4)}$$

Using eq. (6.3) and eq. (6.4), the single scattering fraction can be recovered as

$$\text{Single scattering fraction} = (S_{max} - S_{min}) / (S_{max} + S_{min}) \qquad \text{eq. (6.5)}$$

Using eq. (6.5), the fraction of single scattering is extracted for 2 μm of various concentrations: ($\Phi = 1 \times 10^{-5}$ ($A = 13\%$), $\Phi = 5 \times 10^{-5}$ ($A = 31\%$), $\Phi = 1 \times 10^{-4}$ ($A = 48\%$), $\Phi = 3 \times 10^{-4}$ ($A = 86\%$), $\Phi = 5 \times 10^{-5}$ ($A = 95\%$), and $\Phi = 7 \times 10^{-5}$ ($A = 99\%$) as listed in table 6.1.

| $\Phi$ | $A$ | Single Scattering Fraction |
|---|---|---|
| $1 \times 10^{-5}$ | 14% | 0.96522 |
| $5 \times 10^{-5}$ | 31% | 0.94711 |
| $1 \times 10^{-4}$ | 48% | 0.88093 |
| $3 \times 10^{-4}$ | 86% | 0.62355 |
| $5 \times 10^{-5}$ | 95% | 0.51418 |
| $7 \times 10^{-5}$ | 99% | 0.25111 |

*TABLE. 6.1 Recovered single scattering fraction of 2 μm particles for various concentrations.*

This is in strict connection with the result presented in Chapter 5, where the amplitude *S*(0) of the wave scattered by one particle has been evaluated.

## 6.3. Monte Carlo Simulations

For comparing the fraction of power due to single scattering, Monte Carlo simulation is performed by Dr. Luca Cipelletti (Universite' Montpellier, France) for the present methods geometry.

The Monte Carlo simulation code is described in detail [16]. The simulation consists in tracking the path of a great number of photons passing through the sample. The input parameters are the single scattering differential cross–section and the turbidity of the sample or, equivalently, its transmission.

| *Φ* | *A* | **Turbidity(mm$^{-1}$)** | **Single Scattering Fraction** |
|---|---|---|---|
| 1 x 10$^{-5}$ | 14% | 0.05073 | 0.97613 |
| 5 x 10$^{-5}$ | 31% | 0.13797 | 0.9343 |
| 1 x 10$^{-4}$ | 48% | 0.24077 | 0.88639 |
| 3 x 10$^{-4}$ | 86% | 0.74166 | 0.658 |
| 5 x 10$^{-5}$ | 95% | 1.13245 | 0.4974 |
| 7 x 10$^{-5}$ | 99% | 1.70562 | 0.29253 |

*TABLE. 6.2 Calculated single scattering fraction of 2 µm particles for various concentration.*

The differential scattering cross section is obtained by using the Mie series expansion [3]. The turbidity (-ln (1- *A*) / *L;* *L* is the thickness of the sample) of the samples are calculated by measuring the beam attenuation *A* using calibrated solar cell.

The fractions of single scattering for various concentrations using Monte Carlo simulations are listed in table 6.2.

## 6.4. Extracted Single Fraction Compared with Simulation.

The experimentally extracted and simulated single scattering fraction is shown in figure 6.3 as function of turbidity.

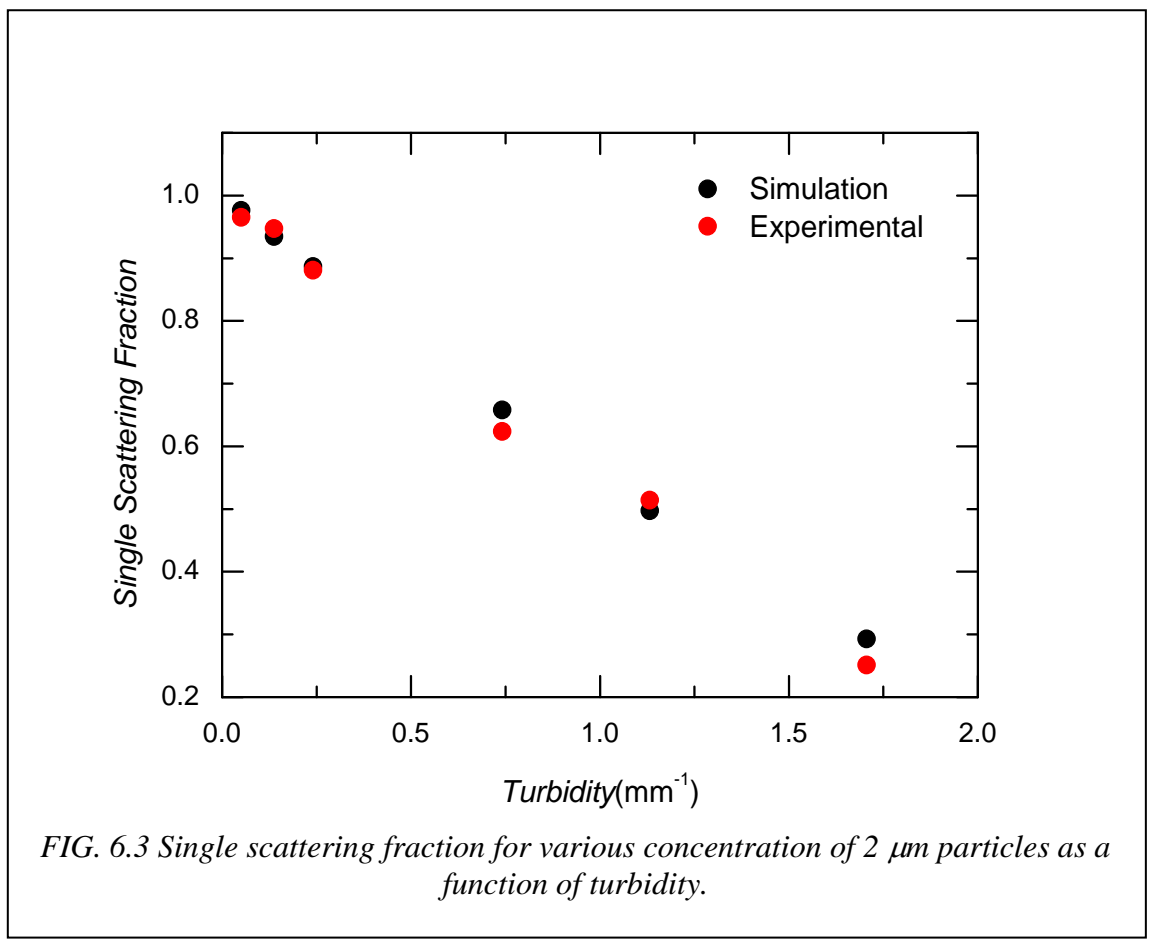

*FIG. 6.3 Single scattering fraction for various concentration of 2 μm particles as a function of turbidity.*

The recovered single scattering fraction using Near Field method shows excellent agreement with the Monte Carlo simulation. Thus the technique provides the valuable information about the total scattered power due to single scattering alone.

## 6.5 Conclusion

The technique also differentiates the single scattering and multiple scattering contribution (the fast dynamics is due to multiple scattering) in the dynamic analysis described in chapter – 2. The fast dynamics is due to multiple scattering and the slow dynamics is due to convective motions (Chapter 4).

The Near Field scattering method allows to unambiguously identify the fraction of the total scattered power that is due to single scattering alone, thus allowing the application of the method even when multiple scattering is rampant (beam attenuation up to 99%).

This allowed to extract the information about the 2-D scattered amplitude $S(0)$, to verify the Optical Theorem for the smallest diameter samples.

*Chapter 7*

*Kinetics of Colloidal Aggregation*

# Chapter 7

# Kinetics of Colloidal Aggregation

Optical thickness is a quantity of interest in phase transitions of colloidal studies such as colloidal aggregation etc. To study this quantity, the knowledge of phase delay of scattered wave is necessary and this information can be obtainable with Near Field based low angle scattering.

In this chapter, the study on aggregating colloids using the Near Field based low angle scattering technique is presented.

## 7.1 Phase Delay Measurement on Aggregating Colloids

The colloids used are 0.3 μm in diameter of polystyrene particles (Duke Scientific Corporation) suspended in water / heavy water. The volume fraction used is $\Phi = 10^{-4}$ and the aggregation is induced by adding the divalent salt Magnesium chloride of 40 mM.

Under these conditions, the aggregation is rather slowly (several hours) following the modality of a reaction limited colloidal aggregation (RLCA). The mass fractal dimension accepted for RLCA is expected to be $D_m = 2.1$ [17].

In this study, the effective sample –sensor distance $z$ used is 83.6 cm and the sensor used is 8-bit Jai CV-M50 industrial monochrome CCD sensor. The dimension of the single pixel is 8.6 μm. All others are same as discussed in Chapter 2.

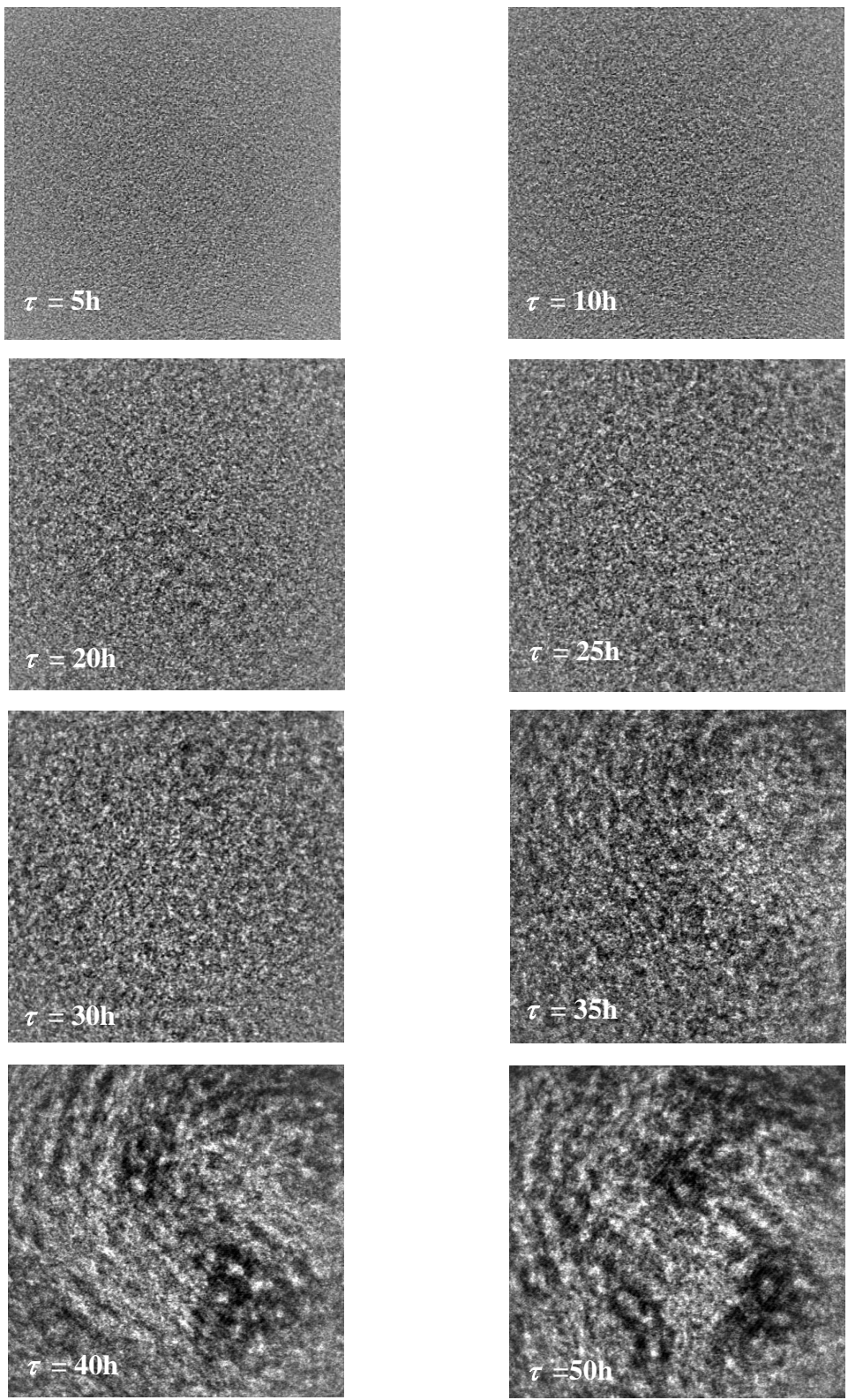

*FIG. 7.1 Differential speckle images of aggregating colloids at different times.*

A set of $N$ near field speckle distribution of aggregating colloids are recorded continuously with a delay time $\tau = 10$ s. The recorded heterodyne signals are analysed using differential double frame analysis (Chapter 3).

The differential speckle images of aggregating colloids are shown in figure 7.1 at different times. The recovered spatial power spectra of aggregating colloids at different times are shown in figure 7.2.

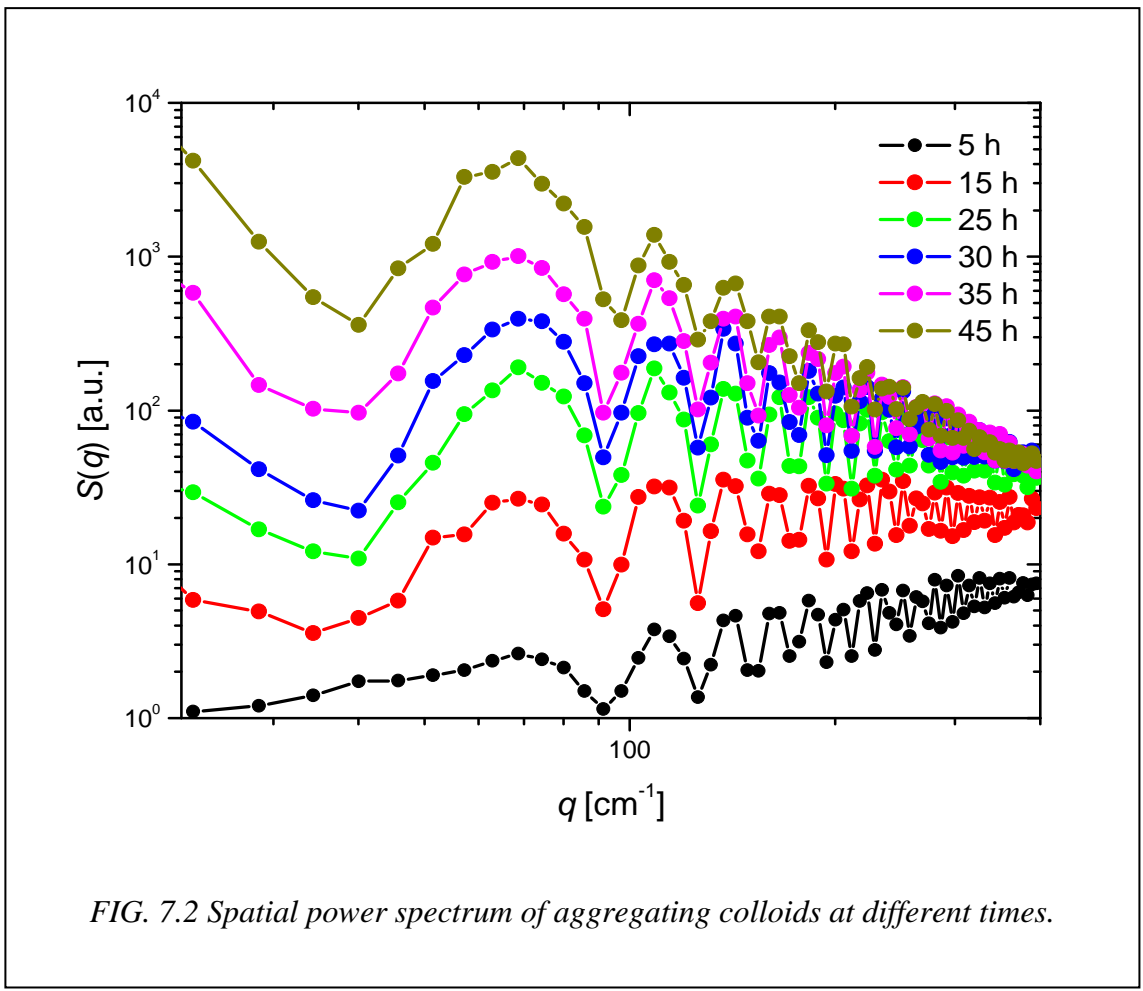

*FIG. 7.2 Spatial power spectrum of aggregating colloids at different times.*

The figure 7.2 shows the typical behaviour of colloidal aggregation with a strong increase ($< 3$ decades) in the zero-$q$ scattered intensity with the roll-off moving to small $q$ and the large $q$ data lying on the same asymptote. Later is the signature of fractal aggregate morphology, representing the mass fractal dimension $D_m$ recovered by fitting the power spectrum using the so called Fisher-Burford (F-B) function [18] as given by

$$S(q) = \frac{S(q=0)}{[1+(2/3D_m)R_G^2 q^2]^{D_m/2}} \qquad \text{eq. (7.1)}$$

The fitting is shown in figure 7.3. The fitting parameters are the zero-$q$ scattered intensity, the fractal dimension $D_m$ and the cluster radius of gyration $R_G$. The fitting reported is satisfactory with the data containing oscillations due to phase correlations and allowed to estimate $D_m = 2.1 \pm 0.1$ as expected for RLCA and $R_G \sim 1$ mm.

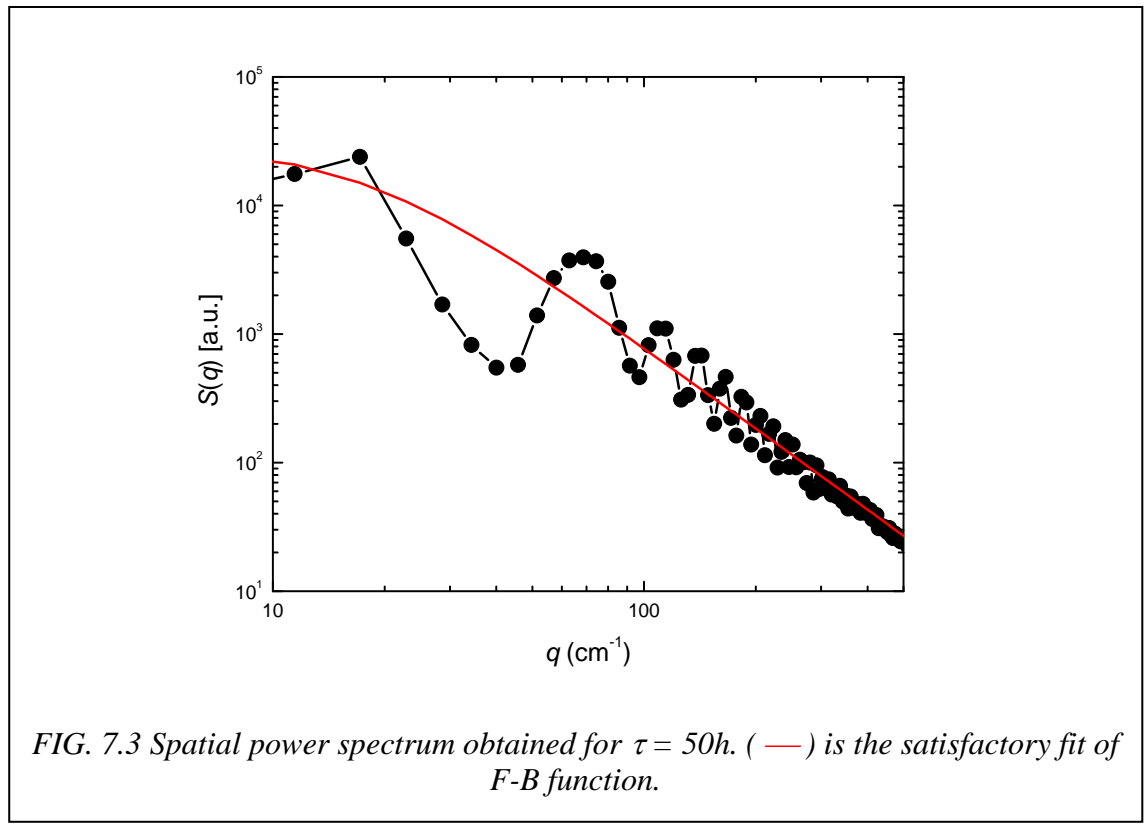

*FIG. 7.3 Spatial power spectrum obtained for $\tau = 50h$. ( —— ) is the satisfactory fit of F-B function.*

As discussed in Chapter 4 & 5, the phase delay between the incident and the scattered wave can be recovered by analysing the Talbot-like oscillations in the obtained spatial power spectra at different times.

The figure 7.4 shows the measured phase delays of the aggregating colloids at different times.

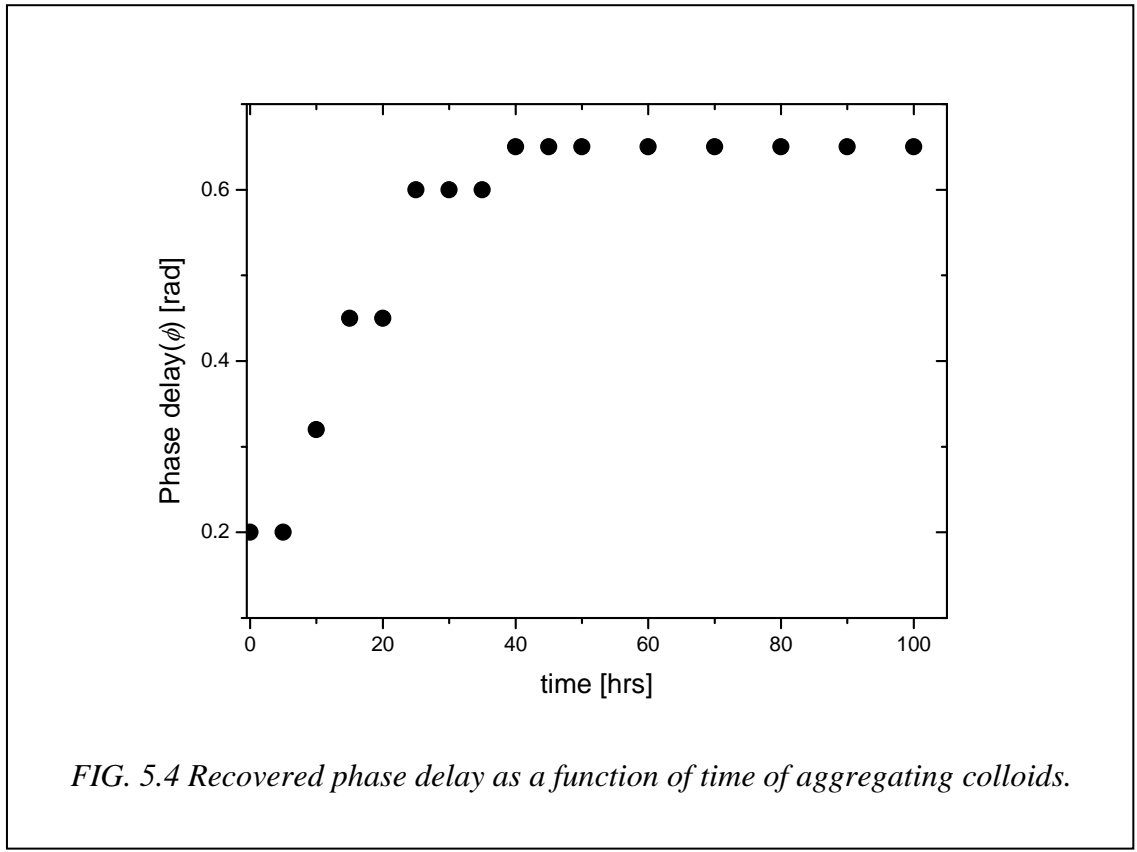

*FIG. 5.4 Recovered phase delay as a function of time of aggregating colloids.*

## 7.2 Calculation of Phase Delay for Aggregating Colloids

As analysed in detail (Chapter 5), for a given experimental apparatus, and for a fixed sample-sensor distance, the positions of the Talbot-like oscillations are determined by the phase lag suffered by the spherical wave emitted by each scatterer with respect to the transmitted.

This lag can be evaluated on the basis of the Mie theory, but under the assumption of Rayleigh-Gans scattering conditions things can be cast in a simpler way [3]. This assumption is valid whenever the optical depth of each scatterer (that is the refractive index multiplied by the path length of light within the scatterer) is much smaller than the wavelength of light. If $a$ is the scatterer radius, $k$ the wavevector, $x = k\,a$ and m the relative refractive index, Rayleigh-Gans conditions are fulfilled until

$$m - 1 \ll 1; \qquad 2\,k\,a\,(m - 1) \ll 1 \qquad \text{eq. (7.2)}$$

The expression for the forward scattered wave can then be reduced to the simple form

$$S(0) = x^2\,K(i\,\rho) \qquad \text{eq. (7.3)}$$

where $\rho = 2\,x\,(m - 1)$ and $K(\rho) = \tfrac{1}{2} + e^{-\rho}/\rho + (e^{-\rho} - 1)/\rho^2$. From this relation the phase lag can be easily obtained. For example, it can be easily obtained from a plot of the real and imaginary part of the complex number $x^{-2}\,S(0)$ (the running parameter is $\rho$). The phase lag is then given by $\operatorname{Im}[x^{-2}S(0)] / \operatorname{Re}[x^{-2}\,S(0)]$.

During the aggregation process the size of the aggregates grow up, while the number density of the monomers decreases, due to the fractal dimension of the aggregates. This also means that the refractive index decreases as well. Within an aggregate of gyration radius $R_G$, the number of monomers can be estimated as

$$N = (R_G / a)^{D_m} \qquad \text{eq. (7.4)}$$

where $D_m$ is the fractal dimension of the aggregate and a indicates the monomer radius. As a result, the number density of monomers within this aggregate will be simply given by

$$n = N / V = (R / a)^{D_m}\,3/(4\pi)\,R^{-3} \qquad \text{eq. (7.5)}$$

On the basis of the Lorentz-Lorenz formula, if $\mu$ indicates the refractive index of the aggregate

$$3\frac{(\mu^2-1)}{(\mu^2+2)} = 4\pi N_{mol}\alpha \qquad \text{eq. (7.6)}$$

where $N_{mol}$ is the number density of molecules with polarizability $\alpha$ within the aggregate. Assuming that $(\mu -1) \ll 1$ as it is in our case, one finds

$$\mu - 1 = 2\pi N_{mol}\alpha \qquad \text{eq. (7.7)}$$

For an aggregate of a given size, the parameter $\rho$ can be evaluated as

$$\rho = 2\,k\,R_G\,(\mu - 1) = 4\,k\,R_G\,\pi\,N_{mol}\,\alpha \qquad \text{eq. (7.8)}$$

Notice that $N_{mol}$ is given by the number density of monomers within the aggregate $n$, times the number of molecules for each monomer, that is given by $V_0 N_0 = V_0\,\rho_m\,N_A\,/\,A$, where $V_0$ is the volume of a monomer, $N_0$ the number density of the molecules in a monomer, $\rho_m$ is the mass density of a monomer, $A$ its molecular weight and $N_A$ the Avogadro's number. Then

$$\rho = 4\,k\,R_G\,\pi\,n\,V_0\,N_0\,\alpha = 4\,k\,\pi\,(R_G/a)^{Dm}\,R_G^{-2}\,a^3\,N_0\,\alpha \qquad \text{eq. (7.9)}$$

*Case 1: Monomers in the Rayleigh-Gans regime*

The parameter $\rho_0$ for a monomer in the Rayleigh-Gans regime can be written by using the Lorentz-Lorenz relation for the monomer

$$\rho_0 = 2\,k\,a\,(m - 1) = 4\,\pi\,k\,a\,N_0\,\alpha \qquad \text{eq. (7.10)}$$

Finally, the ratio of the parameter $\rho$ at a given time during the aggregation and $\rho_0$ at the beginning can be cast in the following simple form

$$\rho/\rho_0 = (R_G/a)^{Dm-2} \qquad \text{eq. (7.11)}$$

Since the value for $\phi_0$ can be evaluated from the known refractive index $m$, this relation provides the value for the parameter $\rho$ at time t, and then the corresponding phase lag can be obtained.

*Case 2: Monomers out of the Rayleigh-Gans regime*

If the monomer size and/or refractive index are large enough that the conditions $2\,k\,a\,(m - 1) \ll 1$ and $m - 1 \ll 1$ are not satisfied, then the expression for $\rho$ can be obtained on the basis of the known refractive index $m$, through the Lorentz-Lorenz relation

$$3\frac{(m^2-1)}{(m^2+2)} = 4\pi N_0 \alpha \qquad \text{eq. (7.12)}$$

where $N_0$ is the number density of molecules within the monomer. The expression for $N_0\,\alpha$ is then

$$N_0 \alpha = \frac{3}{4\pi}\frac{(m^2-1)}{(m^2+2)} \qquad \text{eq. (7.13)}$$

that can be substituted into the expression for $\phi$

$$\rho = 3k\left(\frac{R_G}{a}\right)^{D_m} R_G^2 a^3 \frac{(m^2-1)}{(m^2+2)} \qquad \text{eq. (7.14)}$$

that can be inserted into the expression for $S(0)$ to get the phase lag.

If the monomers radius is $a = 150$ nm; relative refractive index $m = 1.19$; $k = 10^7$ m$^{-1}$; aggregate gyration radius $R_G = 1$ mm; fractal dimension $D_m = 2.1$.

$$\rho_0 = 2\,k\,a\,(m - 1) = 0.57 \qquad \phi = 0.21 \text{ rad} \qquad \text{eq. (7.15)}$$

$$\rho = \rho_0 \, (R_G / a)^{D_m - 2} = 1.37 \qquad \phi = 0.51 \text{ rad} \qquad \text{eq. (7.16)}$$

By using the results obtained for monomers out of the Rayleigh-Gans approximation, one gets

$$\rho = 3k \left( \frac{R_G}{a} \right)^{D_m} R_G^2 a^3 \frac{(m^2 - 1)}{(m^2 + 2)} = 1.37 \qquad \phi = 0.49 \text{ rad} \qquad \text{eq. (7.17)}$$

The phase delays calculated from eq. (7.15) and eq. (7.16) in two cases are comparable. Also the calculated phase delay is quite satisfactory with the experimentally measured phase delay $\phi = 0.60$ rad.

## 7.3 Conclusions

For the first time, the optical thickness of the aggregating colloids is investigated by measuring the phase delay. The experimentally measured phase delay is satisfactory with the calculated one.

# Chapter 8

# Conclusions

# Chapter 8

# Conclusions

The heterodyne ultra low angle scattering method operating in the deep Fresnel region is capable to do that previously considered as impossible, namely measure the properties of zero-angle forward scattered wave. This is really impossible by any classical light scattering methods as the scattering process is conceived as Bragg reflection from a three dimensional random grating. Accordingly, at any angle is the sum of randomly phased contributions, and the phase of the scattered wave is random and no information about the phase of scattered waves at forward angles. By contrast, the present method exploits the so called Raman-Nath scattering from a two dimensional phase grating in which the light scatters equally into positive and negative orders at symmetric angles that are correlated. The three wave interference on the sensor generates deep oscillations in the power spectrum. These oscillations preserve the information about the phase delay.

Having determined the phase delay and the modulus of amplitude function for a series of mono disperse colloidal particles, the absolute total cross sections are derived according to the Optical theorem. Good agreement with Mie theory is found and for the first time the Optical Theorem is verified experimentally in optics.

The oscillations are due to single scattering and the measurements on various concentrations of mono dispersed colloidal particles reveals the depth of the oscillations decrease as the sample attenuation increases as expected. By analyzing the depth of the oscillations, the single scattering fraction is derived and compared with Monte Carlo simulations. Good agreement with simulations is found and the method can be applicable to study high turbid suspensions (99% beam attenuation). The dynamics study also decomposes separately the single scattering fraction from the multiple scattering.

Finally, the method is used to study the aggregating colloids. The derived phase delay is satisfactory with the calculated one.

*References*

# References

[1] K. P. V. Sabareesh, Sidhartha S. Jena and B. V. R. Tata, *AIP Conf. Proc.* **832** (2006) 307.

[2] Sidhartha S. Jena, Hiren M. Joshi, K. P. V. Sabareesh, B. V. R. Tata and T. S. Rao, *Biophy. Jour.* **91** (2006) 2699.

[3] H. C. Van de Hulst, *"Light Scattering by Small Particles"*, (Dover, New York, 1957).

[4] R. G. Newton, *Am. J. Phys.* **44** (1976) 639.

[5] B. J. Berne and R. Pecora, *"Dynamic Light Scattering with applications to Chemistry, Biology and Physics"* (Wiley, New York, 1976).

[6] M. Giglio, M. Carpineti and A. Vailati, *Phys. Rev. Lett.* **85** (2000) 1416.

[7] F. Ferri, D. Magatti, D. Pescini, M. A. C. Potenza and M. Giglio, *Phys. Rev. E* **70** (2004) 041405.

[8] J. W. Goodman, *"Introduction to Fourier Optics"* (McGraw-Hill, Boston, 1996).

[9] M. Giglio, Doriano Brogiol, M. A. C. Potenza and A. Vailati, *Phys. Chem. Phys.* **6** (2004) 1547.

[10] D. Magatti, M. D. Alaimo, M. A. C. Potenza and F. Ferri, *Appl. Phys. Lett.* **92** (2008) 241101.

[11]   M. D. Alaimo, D. Magatti, F. Ferri and M. A. C. Potenza, *Appl. Phys. Lett.* **88** (2006) 191101.

[12]   M. D. Alaimo, M. A. C. Potenza, D. Magatti and F. Ferri, *J. Appl. Phys.* 102 (2007) 073113-1-8.

[13]   M. J. Berg, C. M. Sorensen and A. Chakrabarthi, *J. Opt. Soc. Am. A* **25** (2008) 1504.

[14]   J. W. Goodman, *"Statistical Optics"* (John Wiley & Sons, New York, 1985).

[15]   G. Bryant, H. Schotman and J. C. Thomas, *Part. Part. Syst. Charact.* **15** (1998) 170.

[16]   L. Cipelletti, *Phys. Rev. E* **55** (1997) 7733.

[17]   M. Y. Lin, H. M. Lindsay, D. A. Weitz, R. C. Ball, R. Klein and P. Meakin, *Nature* **339** (1989) 360.

[18]   M. S. Fisher and R. J. Burford, *Phys. Rev.* **156** (1967) 583.